\documentclass[final]{IEEEtran} 
\usepackage{fancyhdr}
\usepackage{epsfig}
\usepackage{threeparttable}
\usepackage{epsf,epsfig}
\usepackage{amsthm}
\usepackage{amsmath}
\usepackage{amssymb}
\usepackage{amsfonts}
\usepackage[noadjust]{cite}
\usepackage{dsfont}
\usepackage{subfigure}
\usepackage{color}
\usepackage{soul}
\usepackage{tabularx}

\newtheorem{lemma}{Lemma}

\newtheorem{proposition}{Proposition}

\def\E{\mathsf{E}}

\def\phi{\varphi}

\def\l{\left}
\def\r{\right}
\def\({\left(}
\def\){\right)}

\def\b0{{\mathbf{0}}}

\newcommand{\var}{\mathsf{var}}

\newcommand{\Pout}{p_{\mathsf{out}}}

\newcommand{\nn}{\nonumber}

\begin{document}	
\title{Renewable Powered Cellular Networks: \\ Energy Field Modeling  and Network Coverage}
\author{Kaibin Huang, Marios Kountouris and Victor O. K. Li \thanks{ K. Huang and V. O. K. Li are with the Dept. of Electrical and Electronic Engineering at The  University of  Hong Kong, Hong Kong. Email: huangkb@eee.hku.hk, vli@eee.hku.hk. M. Kountouris is with the Mathematical and Algorithmic Sciences Lab, Huawei France R\&D, Paris, France. Email: marios.kountouris@huawei.com. Part of this work has been presented at IEEE Intl. Conf. on Comm. Systems (ICCS) in Nov. 2014.  }}

\maketitle

\begin{abstract}
Powering radio access networks using renewables, such as wind and solar power, promises dramatic reduction in the network operation cost and the  network carbon footprints. However, the spatial variation of the energy field can lead to fluctuations in power supplied to the network and thereby affects its coverage. This warrants research on quantifying the aforementioned negative effect and designing countermeasure techniques, motivating the current work. First, a novel energy field model is presented, in which fixed maximum energy intensity $\gamma$ occurs at Poisson distributed locations, called \emph{energy centers}. The intensities fall off from the centers following an exponential decay function of squared distance and the energy intensity at an arbitrary location is given by the decayed intensity from the nearest energy center. The product between the energy center density and the exponential rate of the decay function, denoted as $\psi$, is shown to determine the energy field distribution. Next, the paper considers a cellular downlink network powered by harvesting energy from the energy field and analyzes its network coverage. For the case of harvesters deployed at the same sites as base stations (BSs), as $\gamma$ increases, the mobile outage probability is shown to scale as $(c \gamma^{-\pi\psi}+p)$, where $p$ is the outage probability corresponding to a flat energy field and $c$ is a constant. Subsequently, a simple scheme is proposed for counteracting the energy randomness by spatial averaging. Specifically, distributed harvesters are deployed in clusters and the generated energy from the same cluster is aggregated and then redistributed to BSs. As the cluster size increases, the power supplied to each BS is shown to converge to a constant proportional to the number of harvesters per BS. Several additional issues are investigated in this paper, including regulation of the power transmission loss in energy aggregation and extensions of the energy field model. 

\end{abstract}

\section{Introduction}
The exponential growth of mobile data traffic causes the energy consumption of radio access networks, such as cellular and WiFi networks, to increase rapidly. This not only places heavy burdens on both the electric grid and the environment, but also leads to huge network operation cost. 
A promising solution for energy conservation is to power the networks using alternative energy sources, which will be a feature of future green telecommunication networks \cite{Pike:RenewableBS:2013}. However, the spatial randomness of renewable energy can severely degrade the performance of large-scale networks, hence it is a fundamental issue to address in network design. Considering a cellular network with renewable powered base stations (BSs), this paper addresses the aforementioned issue by proposing a novel model of the energy field and quantifying the relation between its parameters and network coverage. Furthermore, the proposed  technique of \emph{energy aggregation} is shown to effectively counteract energy spatial randomness. 

Studying large-scale energy harvesting networks provides useful insight to network planning and architecture design. This has motivated researchers to  investigate the effects of both the spatial and temporal randomness of renewables on the coverage of different wireless networks spread over the horizontal plane \cite{Huang:WirelessAdHocNetworkEnergyHarvesting, DhillonAndrews:HetNetEnergyHarvesting, LeeZhangHuang:OppEnergyHarvestingCognitiveRadio:2013, HuangLauArXiv:EnablingWPTinCellularNetworks:2013}. Poisson point processes (PPPs) are used to model transmitters of a mobile ad hoc network (MANET) in \cite{Huang:WirelessAdHocNetworkEnergyHarvesting} and BSs of a heterogeneous cellular network \cite{DhillonAndrews:HetNetEnergyHarvesting}. Energy arrival processes at different transmitters are usually modeled as independent and identically  distributed (i.i.d.) stochastic processes, reducing  the effect of energy temporal randomness to independent on/off probabilities of transmitters. Thereby, under an outage constraint, the conditions on the network parameters, such as transmission power and node density of the MANET \cite{Huang:WirelessAdHocNetworkEnergyHarvesting} and densities of different tiers of BSs \cite{DhillonAndrews:HetNetEnergyHarvesting},  can be analyzed for  a given distribution of energy arrival processes. The assumption of spatially independent energy distributions is  reasonable for specific types of renewables that can power small devices, such as kinetic energy and electromagnetic (EM) radiation, but does not hold for primary sources, namely wind and solar power. To some extent, energy spatial correlation is accounted for in \cite{LeeZhangHuang:OppEnergyHarvestingCognitiveRadio:2013, HuangLauArXiv:EnablingWPTinCellularNetworks:2013} but limited to EM radiation. It is proposed in \cite{LeeZhangHuang:OppEnergyHarvestingCognitiveRadio:2013} that nodes in a cognitive-radio network opportunistically harvest energy from radiations from a primary network, besides intelligent sharing of its spectrum. The idea of deploying dedicated stations for  supplying power wirelessly to energy harvesting mobiles in a cellular network is explored in \cite{HuangLauArXiv:EnablingWPTinCellularNetworks:2013}. In \cite{LeeZhangHuang:OppEnergyHarvestingCognitiveRadio:2013, HuangLauArXiv:EnablingWPTinCellularNetworks:2013}, radiations by transmitters with reliable power supply   form an EM energy field and its spatial correlation is determined by the EM wave  propagation  that does not apply to other types of renewables such as wind and solar power. 

It is worth mentioning that the analysis and design of large-scale wireless networks using stochastic geometry and  geometric random graphs \cite{HaenggiAndrews:StochasticGeometryRandomGraphWirelessNetworks, GuptaKumar:CapWlssNetwk:2000, Boyd:RandomGossip:2006} has been a key research area in wireless networking in the past decade. Similar mathematical tools have also been widely used in the area of \emph{geostatistics} concerning spatial statistics of natural resources including renewables, where research focuses on topics such as model fitting, estimation, and prediction of energy fields \cite{Cressie:SpatialStattistics:1993, Chiles:Geostatistics}. The two areas naturally merge in the new area of large-scale wireless networks with energy harvesting. It is in this largely uncharted area that the current work makes some initial contributions. 

\begin{figure*}[t]
\begin{center}
\subfigure[Base station powered by an on-site harvester.]{\includegraphics[width=8cm]{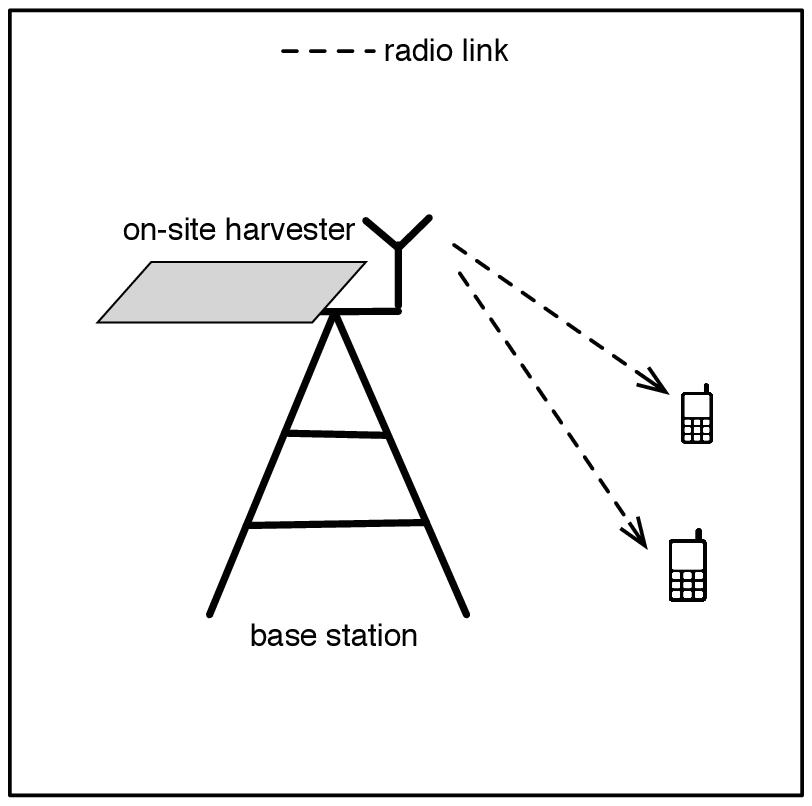}}\hspace{15pt}
\subfigure[Base stations powered by  distributed harvesters.]{\includegraphics[width=8cm]{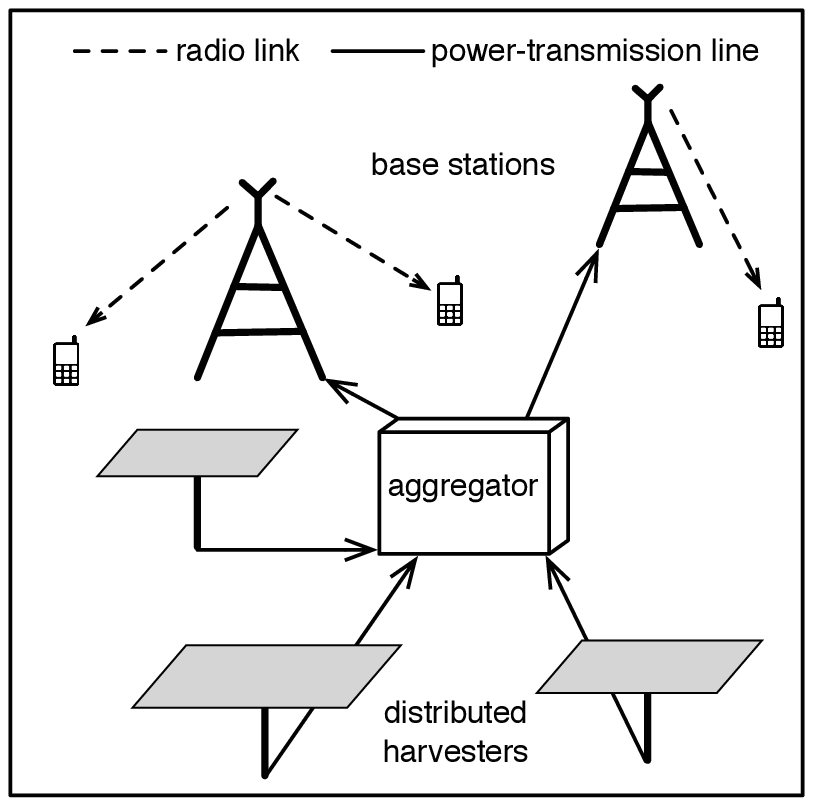}}
\caption{Base stations are powered by energy harvesting using either (a) on-site harvesters or (b) distributed harvesters.}
\label{Fig:Harvesters}
\end{center}
\end{figure*}

Intermittence of renewables introduces stochastic constraints on available transmission power for a wireless device. This calls for revamping classic information and communication theories to account for such constraints and it has recently attracted extensive research efforts \cite{OzelUlukus:InfoTheoEnergyHarvestComm, HoZhang:EnergyAllocatnEnergyHarvestConstraints, OzelUlukus:TransEnergyHarvestFading:OptimalPolicies:2011, Yener:EnergyHarvestingInfChannels:2012,  HuangZhang:EnergyHarvestingRelay:2013, Simeone:MACSensorNetEnergyHarvest:2012, Deniz:EnergyHarvestingBatteryImperfection:2012}. As shown in \cite{OzelUlukus:InfoTheoEnergyHarvestComm} from an  information-theoretic perspective,  it is possible to avoid capacity loss  for an AWGN channel due to energy harvesting provided that a battery with infinite capacity is deployed to counteract the energy randomness. From the communication-theoretic perspective,  optimal power control algorithms are proposed  in \cite{HoZhang:EnergyAllocatnEnergyHarvestConstraints, OzelUlukus:TransEnergyHarvestFading:OptimalPolicies:2011} for single-user systems with energy harvesting, which adapt to the energy arrival profile and channel state so as to maximize the system throughput. These approaches have been extended to design more complex energy harvesting systems, such as interference channels \cite{Yener:EnergyHarvestingInfChannels:2012} and relay channels \cite{HuangZhang:EnergyHarvestingRelay:2013}, to design medium access protocols \cite{Simeone:MACSensorNetEnergyHarvest:2012}, and to account for practical factors such as non-ideal batteries \cite{Deniz:EnergyHarvestingBatteryImperfection:2012}. The prior work assumes fast varying renewable energy sources such as kinetic activities and EM radiation for which adaptive transmission proves an effective way for coping with energy temporal randomness. In contrast, alternative sources targeted in this paper, namely  wind and solar radiation,  may remain static for minutes to hours, which are orders of magnitude longer than the  typical  packet length of  milliseconds. In other words, these sources are characterized by a high level of spatial randomness but very gradual temporal variations. Thus,  this paper focuses on counteracting energy spatial randomness instead of adaptive transmission that is ineffective for this purpose.

In practice, spatial renewable energy maps   are created   by interpolating data collected from sparse measurement stations separated by distances typically of tens to hundreds of kilometers \cite{NRELWebSite, Sen:SpatialInterpolationSoloarIrradiation:2001, Goodin:InterpolationWindField:1979}. The measurement data is too coarse for constructing the energy field targeting next-generation small cell networks with cell radiuses  as small as  tens of meters.  Furthermore, the field depends not only on atmospherical conditions (e.g., cloud formation and mobility) but also on the harvester/BS deployment environment such as locations and heights  of buildings, trees and cell sites. Mapping the energy field  requires complex datasets  that are difficult to obtain. Even if such datasets are available, the resultant   energy-field model may not allow  tractable network performance analysis. This motivates the current approach of modeling energy fields using spatial random processes derived from PPPs. Such processes with a well developed theory   are widely agreed to be suitable models for natural phenomena  and resources \cite{Kingman93:PoissonProc, Cressie:SpatialStattistics:1993} as well as random wireless networks \cite{HaenggiAndrews:StochasticGeometryRandomGraphWirelessNetworks} and thus provide a desirable tradeoff between tractability and practicality. Note that  stochastic-geometry network models in the  literature were developed to overcome the same difficulty of lacking  real data \cite{HaenggiAndrews:StochasticGeometryRandomGraphWirelessNetworks}.  In the proposed model, the random locations of fixed maximum energy intensity (denoted as $\gamma$), called \emph{energy centers}, are modeled as a PPP with density $\lambda_e$, corresponding to sites  on top  of tall buildings with exposure to direct sunshine or strong winds.  From an energy center, the energy intensity decays \emph{exponentially} with the squared distance normalized by a constant $\nu$, called the \emph{shape parameter}, specifying the area of influence  by the said center. It is worth mentioning that the decay function is popularly used in solar-field mapping \cite{Barnes:NumericalWeatherMap:1964} and atmospheric mapping \cite{Sen:SolarEnergy:2004}. The energy intensity at an arbitrary location is then given by the decayed intensity with respect to the nearest energy center.  The  \emph{characteristic parameter} $\psi$ of the energy field, defined as   $\psi = \lambda_e\nu$, is shown later to determine its  distribution.  It is worth mentioning that general insights  obtained in  this work using the above model, such as that  increasing energy spatial correlation reduces outage probability,  also hold for other  models so long as they observe  the basic principle of renewable energy field: energy spatial correlation between two locations increases as their separation distance decreases  and vice versa  \cite{Sen:SpatialInterpolationSoloarIrradiation:2001}.

In the paper, BSs of the cellular network are assumed to be deployed on a hexagonal lattice while mobiles are distributed as a PPP. The network is assumed to operate in the noise-limited regime where interference is suppressed using techniques such as orthogonal multiple access or multi-cell cooperation. The regime is  most interesting from the  energy harvesting perspective since network performance is sensitive to variations of  transmission powers or equivalently, harvested energy. For instance, it has been shown that in the interference limited regime, the network coverage is independent of BS transmission power since it varnishes in the expression for the signal-to-interference-and-noise ratio with noise removed  \cite{Andrews:TractableApproachCoverageCellular:2010}.  A mobile is said to be under (network) coverage if an outage constraint is satisfied and the outage probability is the performance metric. Each BS allocates transmission power simultaneously to mobiles, either by equal division of the available power, called \emph{channel-independent transmission}, or by channel inversion, called \emph{channel inversion transmission}. As illustrated in Fig.~\ref{Fig:Harvesters}, each BS is powered by either an \emph{on-site harvester} or a remote (energy) \emph{aggregator} that collects energy generated by a cluster of nearby \emph{distributed harvesters} over transmission lines. Aggregators and distributed harvesters are deployed on  hexagonal lattices with densities  $\lambda_a$ and $\lambda_h$, respectively. Connecting harvesters to their nearest aggregators form harvester clusters. Note that energy aggregation is based on the same principle of tackling energy randomness by energy sharing as other existing techniques designed for renewable powered cellular systems (see e.g., \cite{ChiaZhang:EnergyCooperationRnewableBS:2015, GuoZhang:EnergyCooperationCellular:2014}).  Prior work focuses on algorithmic design while the current analysis targets  network performance.

The main contribution of this work is the establishment of a new approach for designing large-scale renewable powered wireless networks that models the energy field using stochastic geometry and applies such a model to   network design and performance analysis. The analysis based on the approach allows  the interplay of different mathematical tools such as stochastic geometry theory, probabilistic inequalities  and large deviation theory, leading  to the following new  findings. 

\begin{itemize}
\item Consider the on-site harvester case.  If the characteristic parameter $\psi$ is small ($\psi \leq 1/\pi$) and the maximum energy density $\gamma$ is large ($\gamma \rightarrow \infty$), the outage probability monotonically decreases with increasing $\psi$ and $\gamma$ in the form of $(c \gamma^{-\pi\psi}+p)$ with $p$ being the probability corresponding to a flat energy field and $c$ a constant. The result holds for both channel inversion  and channel-independent transmissions. Despite this similarity, the former outperforms the latter by adapting transmission power to mobiles' channels. 

\item Next, consider distributed harvesters and define the size of a harvester cluster as $\lambda_h/\lambda_a$. As the cluster size increases, energy aggregation is shown to counteract the spatial randomness of the energy field and thereby stabilize the power supply for BSs. Specifically, the power it distributes to each BS converges to a constant proportional to the number of harvesters per BS. In other words, the energy field becomes a reliable power supply for the network. 

\item However, an insufficiently high voltage used by harvesters for power transmission to aggregators can incur significant energy loss. It is found  that the loss can be regulated by increasing the voltage inversely with the aggregator density to the power of $\frac{1}{4}$.   

\item Last, key results are extended to two variations of the energy field model characterized by a shot noise process and a power-law energy decay function, respectively. 

\end{itemize}

The remainder of the paper is organized as follows. The mathematical models and metrics are described in Section~\ref{Section:Model}. The energy field model is proposed and its properties characterized  in Section~\ref{Section:Field}. The network coverage is analyzed for the cases of on-site and distributed harvesters in Sections~\ref{Section:Coverage:OnHar} and \ref{Section:Coverage:DistHar}, respectively. The extensions of the energy field model are  studied in Section~\ref{Section:Ext}. Simulation results are presented in Section~\ref{Section:Sim} followed by concluding remarks in Section~\ref{Section:Conclusion}.

\section{Models and Metrics}\label{Section:Model}
The spatial models for the  energy field, energy harvesters, and cellular network are described in the subsections. The notations are summarized in Table~\ref{Tab:Notation}.

\subsection{ Energy Harvester Model}
\emph{Aggregation loss} arises in the scenario of distributed harvesters, corresponding  to the power loss due to transmissions from harvesters to aggregators over cables (see Fig.~\ref{Fig:Geometry}).  It is impractical to assume high voltage transmission from distributed harvesters having small form factors and thus aggregation loss can be significant, which is analyzed in the sequel. On the other hand, each aggregator supplies power to $\lambda_b/\lambda_a$ BSs also over cables,  assuming that  $\lambda_b/\lambda_a$ is an integer for simplicity. Aggregators are assumed to be much larger than harvesters and thus can afford having relatively large transformers. Thus,  high voltages can be reasonably  assumed for power distribution  such that the power loss is negligible. Consequently, the specific graph of connections between aggregators and BSs  has no effect on the analysis except for the number of BSs each aggregator supports. 

The energy field is represented by $\Psi$, which is determined by the function $g(X)$ mapping a location $X\in\mathds{R}^2$ to energy intensity. The discussion of the stochastic geometry  model of the energy field  is postponed to Section~\ref{Section:Field}, whereas its relation with energy harvesting is described below.  Let $g(X, t)$ represent the time-varying version of $g(X)$ with $t$ denoting time. Harvesters are assumed to be homogeneous and time is partitioned into slots of unit duration.  Let $\eta \in (0, 1)$ denote a constant combining  factors such as the harvester physical configuration and  conversion efficiency, referred to as the \emph{harvester aperture} by analogy with the antenna aperture.  Then the amount of energy harvested at $X$ in the $n$-th slot is $\eta g_n(X)$ or equivalently $\eta\int_{n}^{n+1} g(X, t) dt$.  

\begin{table}[t]
\setlength\extrarowheight{3pt}
\caption{Summary of Notations}
\begin{center}
\begin{tabular}{|c|p{6.1cm}|}
\hline
{\bf Symbol}  &   {\bf Meaning}\tabularnewline 
\hline
$\Phi_e, \lambda_e$ & Process of energy centers and its density.\\
$\Phi_h$, $\lambda_h$ & Set of harvesters on a hexagonal lattice and its density.\\
$\Phi_b$, $\lambda_b$ & Set of BSs on a hexagonal lattice  and its density. \\
$\Phi_a$, $\lambda_a$ & Set of aggregators  on a hexagonal lattice  and its density. \\
$f(d)$ & Energy decay function of distance $d$. \\
$g(X)$ & Energy intensity at location $X$.\\
$\gamma$ & Maximum energy intensity of the energy field. \\
$\psi$ & Characteristic parameter of the energy field. \\
$\eta$ & Harvester aperture.\\
$B_0$, $P_0$ & Typical BS and  its transmission power. \\
$C_0$, $K_0$ & Cell served by $B_0$ and the number of mobiles in the cell. \\
$R_{0, n}$, $H_{0, n}$ & Propagation distance and channel coefficient for the $n$-th mobile in $C_0$. \\
$\Pout$, $\epsilon$ & Outage probability and its constraint. \\
$\Delta P_0$ & Reduction of $P_0$ due to aggregation loss. \\
$V$ & Harvester voltage for power transmission. \\
$\tau$ & Fixed multiplier of $P_0$ in $(0, 1)$ representing regulated aggregation loss. \vspace{3pt}\\
\hline
\end{tabular}
\end{center}
\label{Tab:Notation}
\end{table}

\subsection{Cellular Network Model}

\begin{figure}[t]
\begin{center}
\subfigure[On-site harvesters co-located  with BSs.]{\fbox{\includegraphics[width=7cm]{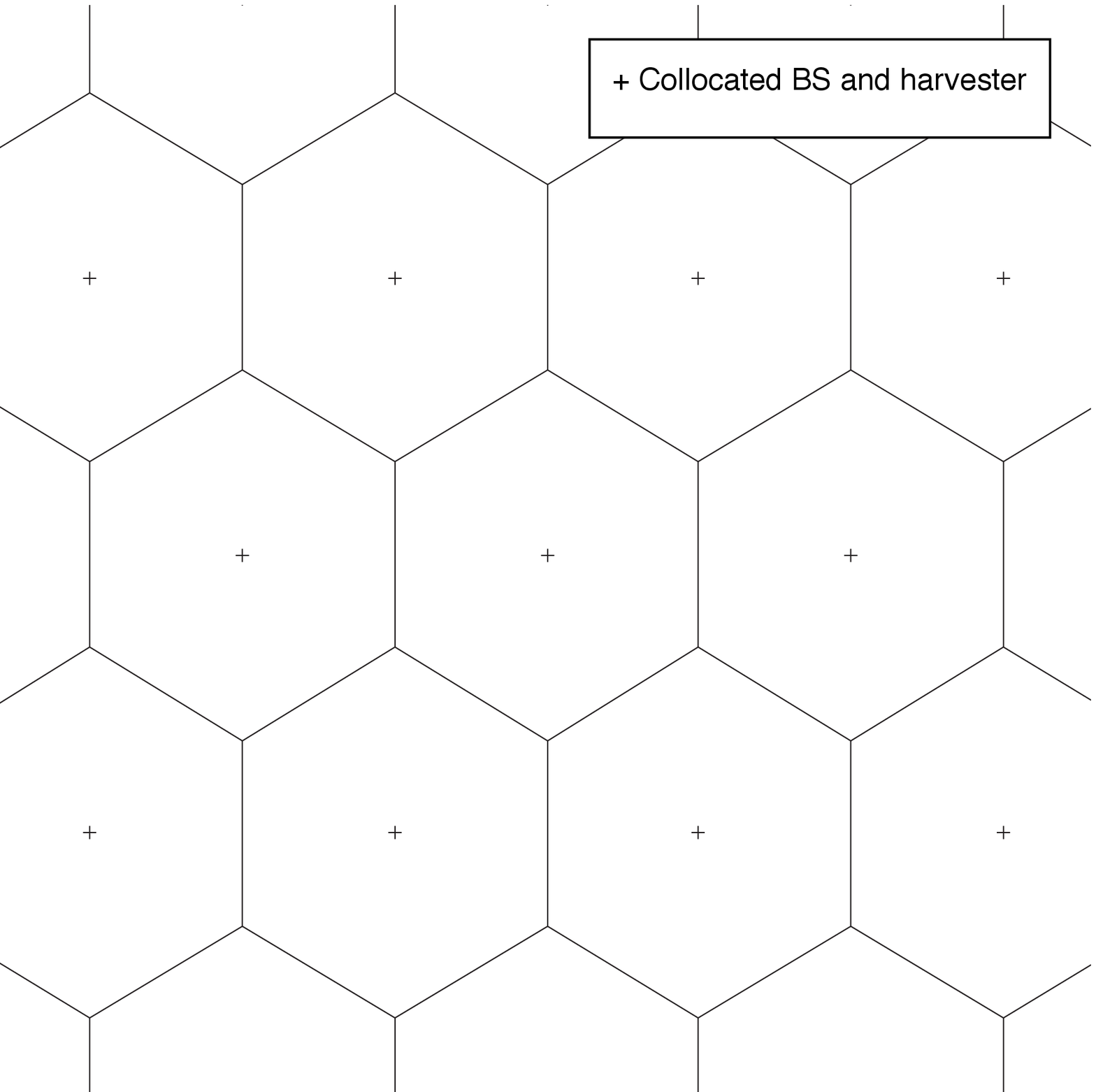}}}\vspace{10pt}
\subfigure[Distributed harvesters.]{\fbox{\includegraphics[width=7cm]{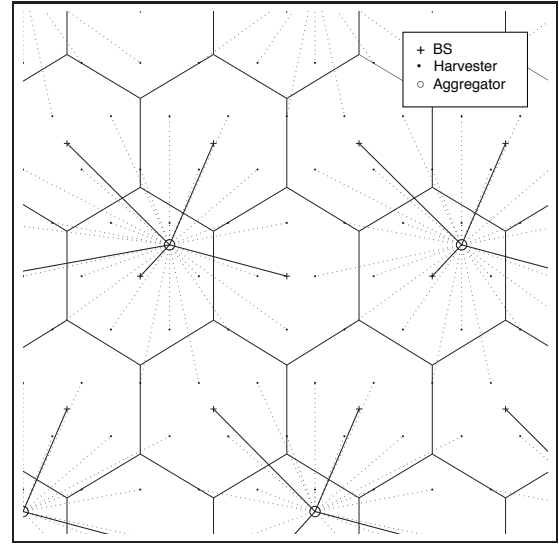}}}
\caption{Geometric patterns  of hexagonal cells, BSs, harvesters and aggregators plotted/marked using solid lines, crosses, dots and circles, respectively.  Power transmission  lines from harvesters to aggregators and those from aggregators to BSs are plotted using dashed and solid lines, respectively.}
\label{Fig:Geometry}
\end{center}
\end{figure}

As illustrated in Fig.~\ref{Fig:Geometry}, the traditional  model of a cellular network is adopted, in which the BSs are deployed on a hexagonal lattice with density $\lambda_b$, denoted as $\Phi_b$,  and consequently the plane is partitioned into hexagonal cells with areas of $1/\lambda_b$. Let $B_0$, $P_0$ and $K_0$ denote 
the typical BS, its transmission power and the number of  simultaneous mobiles  the BS serves, respectively.  Since adaptive transmission is effective for counteracting energy spatial randomness as mentioned earlier, $P_0$ is assumed to be fixed and equal to the power supplied to the BS so as to keep the exposition concise. In addition,  circuit power consumption of each BS is assumed to be negligible compared with $P_0$.\footnote{The analysis can be  modified to account for fixed BS circuit power, denoted as $\phi$,  by replacing $P_0$ with $(P_0 - \phi)$ in the definitions of outage probability in \eqref{Eq:Pout:CI} and \eqref{Eq:Pout:UA} in the sequel.  However, the notation and derivation steps are made tedious while the analytical methods and key insights remain unchanged.} Mobiles are assumed to be distributed as a PPP with density $\lambda_u$ and all  mobiles are scheduled for simultaneous transmissions. Then $K_0$ is a Poisson random variable with mean $\lambda_u/\lambda_b$. A  signal transmitted by a BS at $X\in\mathds{R}^2$ with power $P$ is received at a mobile at $Y\in\mathds{R}^2$ with power given as $PH_{XY} |X-Y|^{-\alpha}$, where $|X-Y|$ is the Euclidean distance between $X$ and $Y$, $\alpha > 2$  the path-loss exponent and the random variable $H_{XY}$, called a channel coefficient,  models fading or shadowing. The channel coefficients are assumed to be  i.i.d. Assuming unit noise variance, the received power also gives the received SNR. In the noise-limited network,  the condition for reliable data decoding at a mobile is specified by the outage constraint that the received SNR exceeds a given threshold $\theta$ except for a small probability $\epsilon$.

Define an outage event as one that the receive SNR of a typical active mobile is below $\theta$. The outage probability is the  network performance metric and defined for the two transmission strategies as  follows. First, consider channel-independent transmission, where a BS equally allocates  supplied power for transmission to mobiles. Let  $R_{0}$ and $H_{0}$ denote the propagation distance and the channel coefficient of a typical user in the typical cell, respectively.  Then, the outage probability, denoted as  $\Pout^{\text{id}}$, can be written as 
\begin{align}
\Pout^{\text{id}} &= \Pr\l( \frac{P_0 H_0 R_0^{-\alpha}}{K_0}   < \theta\r)  \label{Eq:Pout:CI}
\end{align}
where given equal power allocation, the transmission power for each of the  $K_0$  users in the typical  cell is $P_0/K_0$. Alternatively, the outage probability can be defined with respect to a set of users in the same cell as the probability that the received SNR for least one user is below the threshold. Following similar steps as in the subsequent  analysis, 
one  expect the resultant outage probabilities to have similar  expressions  as those in Propositions~\ref{Prop:OnHar:CITX}-\ref{Prop:OnHar:INVTX:LD} that arise  mainly from  the distribution of the energy field instead of the the specific definition of outage probability. 

Next, consider channel inversion transmission.  Transmission power required for each mobile is computed by channel inversion in an attempt  to ensure that  the receive SNR is equal to $\theta$. Specifically, the $n$-th mobile in the typical cell is under coverage if the allocated transmission power is no smaller than $\theta R_{0, n}^{\alpha}/H_{0, n}$, where $R_{0, n}$ and $H_{0, n}$ represent the corresponding propagation distance and channel coefficient, respectively.  It is difficult to write down the outage probability for a typical active mobile, denoted as $\Pout^{\text{iv}}$,  since it depends on the channels of other active mobiles sharing the BS. However, it can be bounded using the union bound as 
\begin{equation}\label{Eq:Pout:UA}
\Pout^{\text{iv}} \leq \Pr\l(P_0 < \theta\sum_{n=1}^{K_0} \frac{R_{0, n}^{\alpha}}{H_{0, n}}\r).  
\end{equation}
The upper bound in \eqref{Eq:Pout:UA} is the probability of the event that  \emph{at least} one user is in outage, which  is the union of the outage events of individual users.

\section{Energy Field Model and Its Properties}\label{Section:Field}

\subsection{Energy Field Model}
\begin{figure*}
\begin{center}
\subfigure[$\lambda_e = 1, \nu = 0.1$]{\includegraphics[width=8cm]{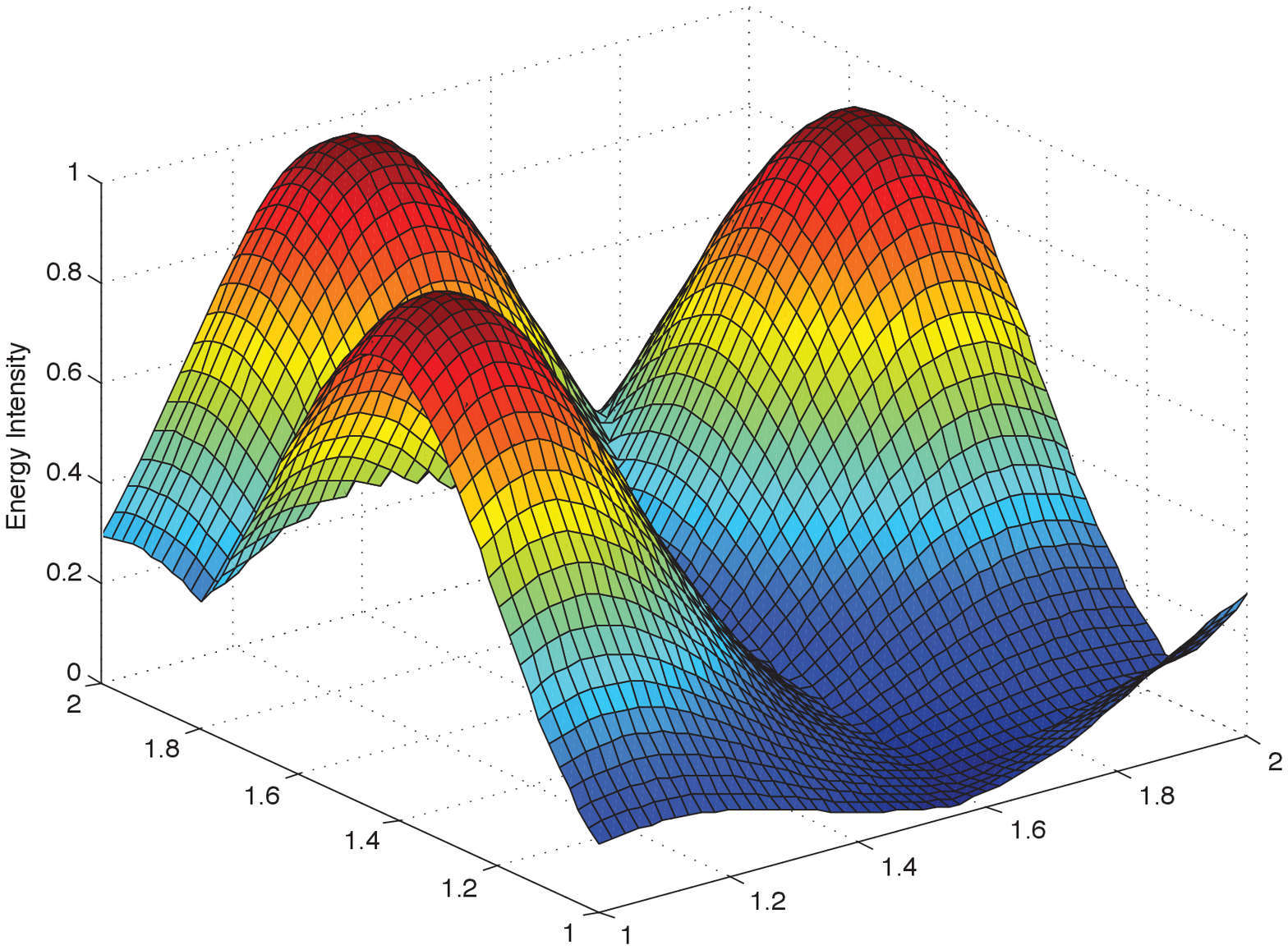}}\hspace{10pt}
\subfigure[$\lambda_e = 1, \nu = 1$]{\includegraphics[width=8cm]{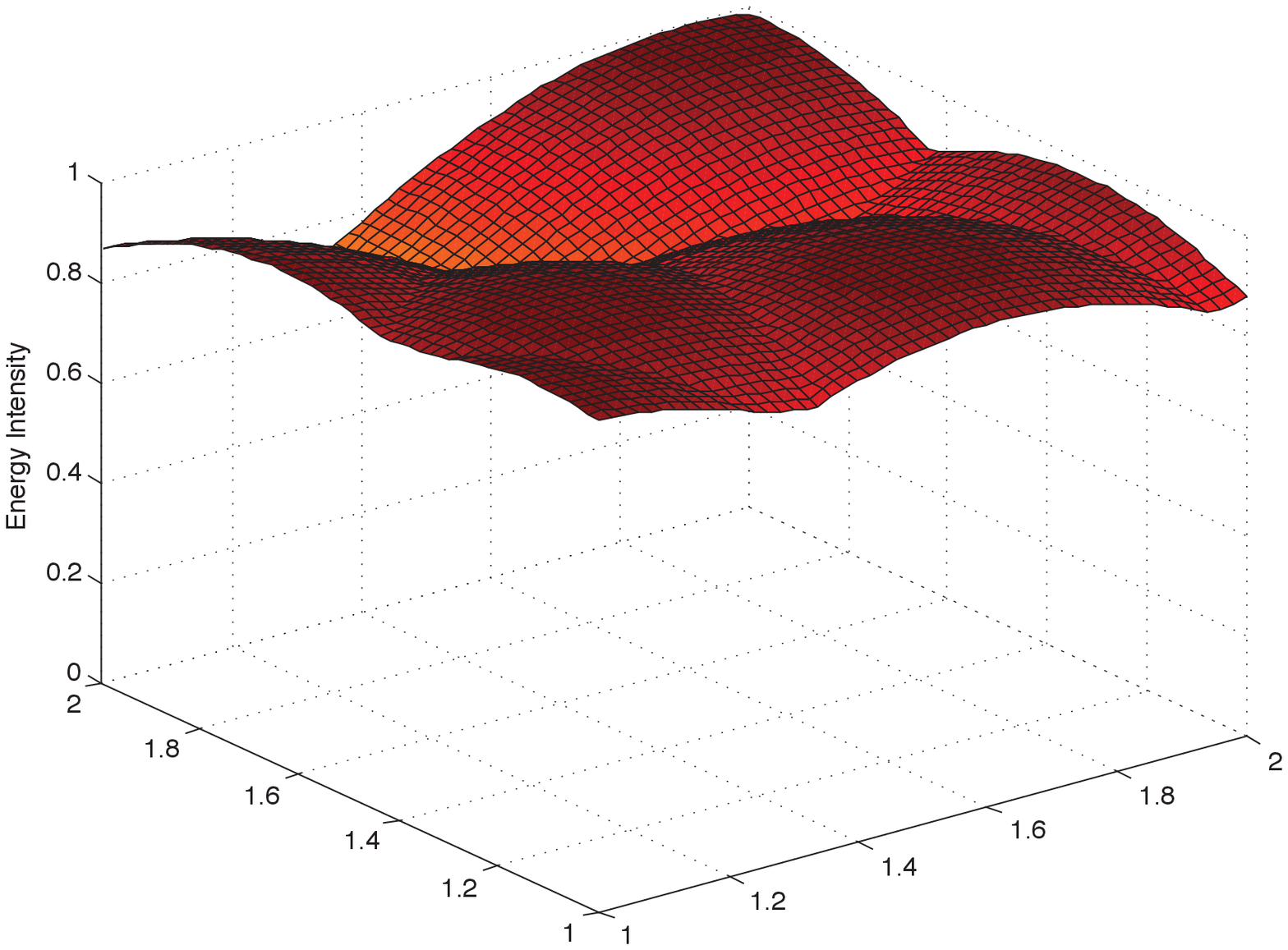}}\\
\subfigure[$\lambda_e = 10, \nu = 0.1$]{\includegraphics[width=8cm]{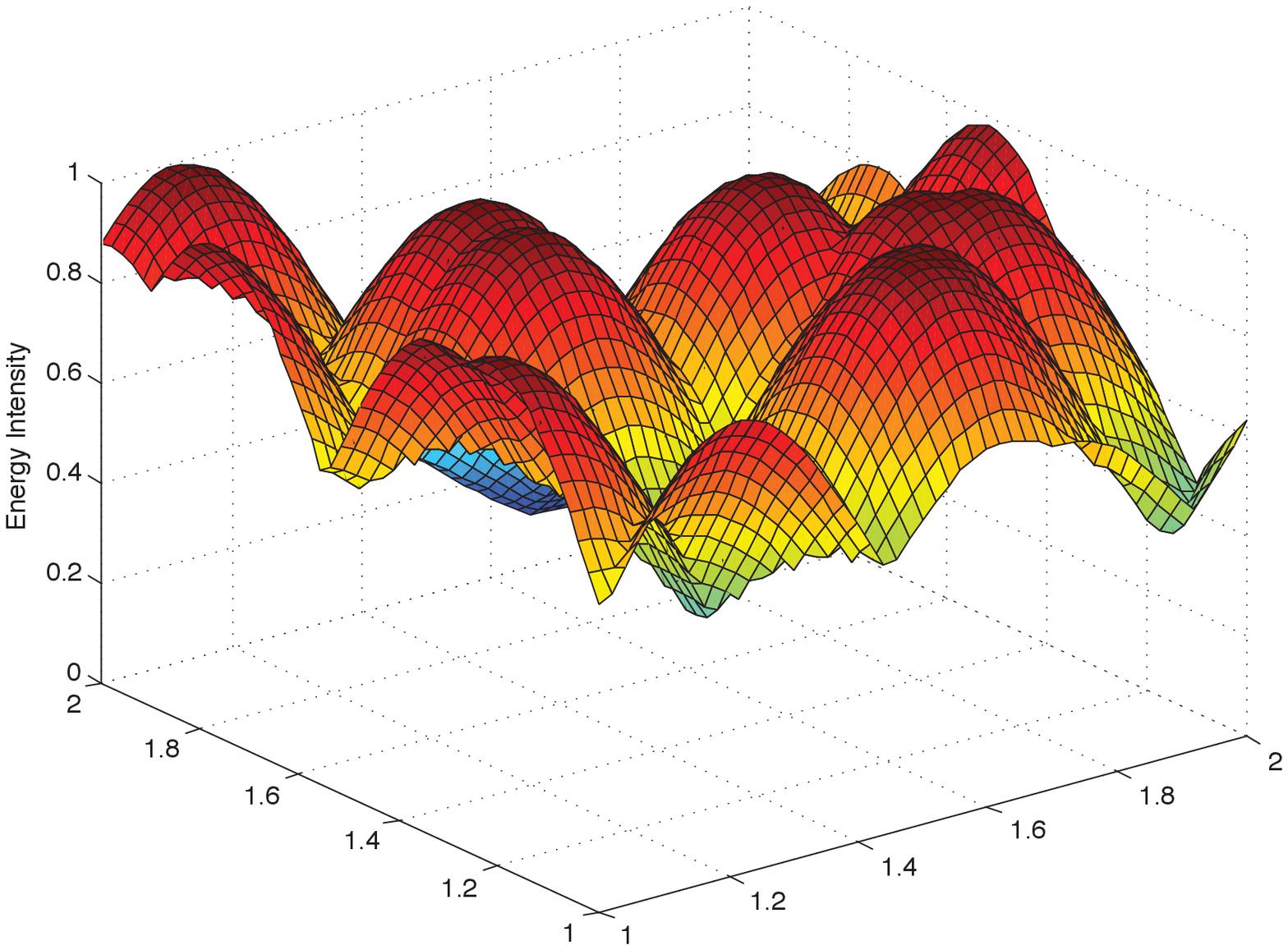}}\hspace{10pt}
\subfigure[$\lambda_e = 10, \nu = 1$]{\includegraphics[width=8cm]{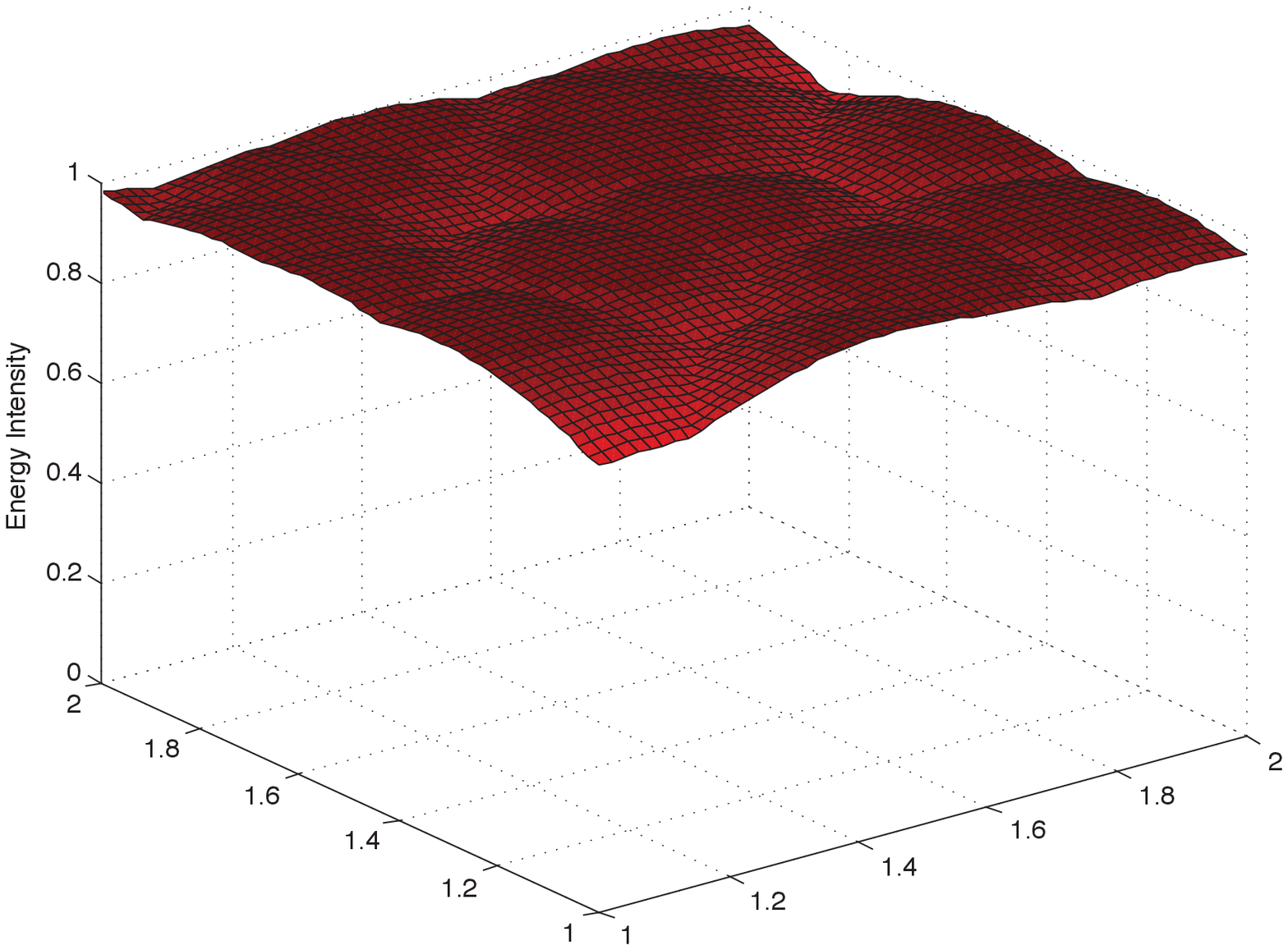}}\\
\caption{Energy field for different combinations of the energy center density $\lambda_e$ and shape parameter $\nu$ and different realizations of the energy-center process. }
\label{Fig:EnergyField}
\end{center}
\end{figure*}

The energy field $\Psi$ refers to the set of energy intensities at different locations in the horizontal plane with the maximum denoted as $\gamma > 0$.  The PPP modeling for the energy centers and its density are denoted as $\Phi_e\subset \mathds{R}^2$ and $\lambda_e$, respectively. The energy intensity function $g(X)$ is defined for a given location $X\in\mathds{R}^2$ as follows: 
\begin{equation}\label{Eq:EnDen}
g(X) =\gamma\max _{Y \in \Phi_e} f(|X-Y|)
\end{equation}
where the energy decay function $f$ is  
\begin{equation}\label{Eq:DecayFun}
f(d) = e^{-d^2/\nu},\qquad d > 0. 
\end{equation}
Note that $g(X)$  is  a \emph{Boolean random function} \cite{Serra:BooleanRandomFun:1989}. 
The positive  parameter $\nu$ in \eqref{Eq:DecayFun} controls the shape of $f$, thus  called the \emph{shape parameter}, and thereby determines the area of influence  of an energy center.   Then the characteristic parameter  $\psi$ is defined as $\psi = \nu\lambda_e$. A large value of $\psi$ corresponds to an energy field with gradual spatial variation and a small value indicates that the  field has  a lot of ``shadows", where energy intensities are much lower than the peak. As observed from the plots in Fig.~\ref{Fig:EnergyField}, increasing $\lambda_e$  or decreasing $\nu$ introduces more ``ripples" in the energy field; the field is almost flat for a large characteristic parameter    e.g.,  $\psi = 10$ (or $\lambda_e = 10$ and $\nu = 1$).  

An alternative model of the energy field can result from replacing the max operator in the energy intensity function in \eqref{Eq:EnDen} with a summation. The two  models are  shown in  Section~\ref{App:Field:Alt} to have similar  stochastic properties in the operational regime of interest. Furthermore,  the energy field based on a different energy decay function is considered   in Section~\ref{Eq:Field:Ext} and its effect on the network coverage is analyzed.

\subsection{Energy Field Properties}
First, the distribution function of the energy intensity can be easily obtained by relating it to a Boolean model. To this end, define $r(x)$ as the  distance from an energy center to a location with  the decayed energy intensity $x$. Then $r(x)$ can be obtained from the equation $f(r(x)) = x$ with $f$  in \eqref{Eq:DecayFun} as follows: 
\begin{equation}
r(x) = f^{-1}\l(\frac{x}{\gamma}\r)= \sqrt{\nu\ln \frac{\gamma}{x}}. \label{Eq:Radius}
\end{equation}
  Moreover, let $B(X, r)$ denote a disk in $\mathds{R}^2$ centered at $X$ and with a radius $r$.  The region of the energy field where energy intensities exceed a threshold $x$ corresponds to  a Boolean model $\bigcup_{X \in \Phi_e} B(X, r(x))$.  Then for  given $X$ and $x \in  [0, \gamma]$, 
the distribution function of $g(X)$ can be written in terms of  the model and obtained as 
\begin{align}
\Pr(g(X) \leq x) &= \Pr\l(X \notin  \bigcup_{X \in \Phi_e} B(X, r(x))\r)\nn\\
& = e^{-\pi \lambda_e r^2(x)}\nn\\
& = \l(\frac{x}{\gamma}\r)^{\pi\psi} \label{Eq:Field:Dist}
\end{align}
where the last equality is obtained by substituting \eqref{Eq:Radius} and using the definition of $\psi$. Next, the mean and variance  of the energy intensity function can be directly  derived using the distribution function in \eqref{Eq:Field:Dist} as follows:
\begin{align}
\E[g(X)] &= \frac{\pi \psi \gamma}{1 + \pi \psi}, \label{Eq:Energy:Mean}\\
\E[g^2(X)] &= \frac{\pi \psi \gamma^2}{2 + \pi \psi}, \nn \\
\var(g(X)) &= \frac{\pi \psi \gamma^2}{(2 + \pi \psi)(1+\pi \psi)^2}. \nn 
\end{align}
As $\psi \rightarrow \infty$,  $\E[g(X)]$ is seen to converge to $\gamma$ while $\var(g(X))$ diminishes inversely with  $\psi^2$. On the other hand, as $\psi \rightarrow 0$, both quantities become  proportional to $\psi$. 

Last, the joint distribution of the energy intensities $g(X_1)$ and $g(X_2)$ at two different locations $X_1$ and $X_2$ is derived.  They are independent if $|X_1 - X_2| \geq r(x_1) + r(x_2)$: 
\begin{align}
\Pr\l(g(X_1)\leq x_1, g(X_2)\leq x_2\r) &= \Pr\l(g(X_1)\leq x_1\r)\times \nn\\
&\qquad  \Pr\l(g(X_2)\leq x_2\r)\nn\\
&= \l(\frac{x_1x_2}{\gamma^2}\r)^{\pi\psi}. \label{Eq:JCDF}
\end{align}
If $|X_1 - X_2| < r(x_1) + r(x_2)$,   
\begin{equation}\label{Eq:JCDF:a}
\begin{aligned}
&\Pr\l(g(X_1)\leq  x_1, g(X_2)\leq x_2\r) \\
&\qquad \qquad = \l(\frac{x_1x_2}{\gamma^2}\r)^{\pi\nu  \lambda_e} e^{\lambda_e |B(X_1, r(x_1)) \cap B(X_2, r(x_2))|} 
\end{aligned}
\end{equation}
where the operator $|\cdot|$ applied on a set gives its measure (area). 
For ease of notation, rewrite  $r(x_1)$ and $r(x_2)$ as $r_1$ and $r_2$ and define $d = |X_1 - X_2|$. The area of the overlapping region between  $B(X_1, r_1)$ and $B(X_2, r_2)$ is known to be given as 
\begin{align}
&|B(X_1, r_1) \cap B(X_2, r_2)| = r_1^2\cos^{-1}\l(\frac{r_1^2 + d^2 - r_2^2}{2 d r_1}\r)+ \nn\\
&  + r_2^2\cos^{-1}\l(\frac{r_2^2 + d^2 - r_1^2}{2 d r_2}\r) -  d\sqrt{r_1^2 - \l(\frac{r_1^2 + d^2 - r_2^2}{2d}\r)^2}. \nn
\end{align}
For the special case of $x_1 = x_2=x$ (thus $r_1 = r_2 = r$),  
\begin{equation}\label{Eq:JCDF:b}
\Pr\l(g(X_1)\leq x, g(X_2)\leq x\r) = \l(\frac{x}{\gamma}\r)^{2\pi\nu  \lambda_e(1 - \Delta(d/r))} 
\end{equation}
where $\Delta(d/r) \geq 0$ and 
\begin{equation}
\Delta(y) = \frac{1}{2\pi}\cos^{-1}\l(\frac{y}{2}\r) - \frac{y}{4\pi} \sqrt{4 - y^2}. \nn
\end{equation}
Comparing \eqref{Eq:JCDF:a} and \eqref{Eq:JCDF:b}, one can see that reducing the distance between two locations increases the joint cumulative distribution function of the corresponding energy intensities as they become more correlated.

\section{Network Coverage  with On-Site Harvesters}\label{Section:Coverage:OnHar}

\subsection{Network Coverage with Channel-Independent Transmission}
To facilitate analysis, the outage probability is decomposed as $\Pout^{\text{id}} = p_b + p_a$ with $p_b$ and $p_a$ defined as 
\begin{align}
p_a &= \Pr\l(\frac{P_0 H_0 R_0^{-\alpha}}{K_0}   < \theta\l| K_0 H_0^{-1} R_0^{\alpha}\leq \frac{\eta\gamma}{\theta} \r.\r)\times\nn\\
&\qquad \Pr\l(K_0 H_0^{-1} R_0^{\alpha}\leq \frac{\eta\gamma}{\theta} \r),\label{Eq:p_a}\\
p_b &= \Pr\l(K_0 H_0^{-1} R_0^{\alpha}> \frac{\eta\gamma}{\theta} \r). \label{Eq:p_b}
\end{align}
The component $p_a$  is the probability that the energy intensity at the typical BS site is so low as to cause an outage event   at the typical mobile even though  the maximum intensity is sufficiently high for avoiding such an  event. The other component $p_b$ represents  the  probability that the maximum  energy intensity at the typical BS is insufficiently high for avoiding an outage event at the typical mobile, which  occurs due to the combined effect of the number of  simultaneous users $K_0$ being too large, the  channel gain being too small, and the propagation distance $R_0$ being too long. In other words, $p_a$ and $p_b$ are the components of  outage probability arising from  the energy field spatial variation and the limit on the maximum harvested power, respectively. The outage probability is characterized by analyzing $p_a$ and $p_b$ separately. 

To this end, it is useful to define a random variable $\bar{R}_0$ as $\bar{R}_0 = \sqrt{\lambda_b} R_0$ that gives  the propagation distance of a mobile uniformly distributed in a cell of unit area. Note that a hexagon of unit area can be  outer  bounded by a disk with the area of  $\frac{2\pi}{3\sqrt{3}}$. Let $D$ denote the distance from the center of the disk to a random point uniformly distributed in the  disk. Then
\begin{align}
\Pr(\bar{R}_0\leq x) & \geq \Pr\l(D \leq x\r)\nn\\
& = \frac{3\sqrt{3}}{2} x^2, \qquad 0 \leq x \leq \sqrt{\frac{2}{3\sqrt{3}}}. \label{Eq:R:Dist:Inner}
\end{align}

The probability $p_a$ is  shown in the following lemma to be an exponential function of the energy field characteristic parameter $\psi$, when $\psi$ is sufficiently small. The proof of the lemma is provided in Appendix~\ref{Lem:Pa:Proof}. 
\begin{lemma}\label{Lem:Pa}The probability $p_a$ can be upper bounded as 
\begin{equation}
p_a  \leq \frac{2}{2+\alpha \pi\psi} \l(\frac{c_2\theta\E\l[H_0^{-1}\r]\lambda_u}{\gamma\eta\lambda_b^{1+\frac{\alpha}{2}}}\r)^{\min(\pi\psi, 1)}
\end{equation}
where $c_2 = \l(\frac{2}{3\sqrt{3}}\r)^{\frac{\alpha}{2}}$. 
\end{lemma}

One can observe from \eqref{Eq:p_b} that $p_b$ is the tail probability of the product of multiple random variables and thus it is difficult to derive a closed-form expression for the probability. Following typical approaches, we apply a  probabilistic inequality, namely Markov's inequality,  to obtain an upper bound on $p_b$ and large deviation theory to characterize its asymptotic scaling.
The following lemma is obtained based on \emph{Markov's inequality} with the proof in Appendix~\ref{Lem:Pb:Markov:Proof}. 
\begin{lemma}\label{Lem:Pb:Markov}\emph{The probability $p_b$ in \eqref{Eq:p_b} can be upper bounded as
\begin{equation}
p_b  \leq \frac{c_3\theta\lambda_u\E\l[H_0^{-1}\r]}{\eta\gamma\lambda_b^{1+\frac{\alpha}{2}}},  \label{Eq:Pa:ub}
\end{equation} 
where the constant $c_3 = \frac{2}{2+\alpha}\l(\frac{2}{3\sqrt{3}}\r)^{\frac{\alpha}{2}}$. 
}
\end{lemma}

Given $\Pout^{\text{id}} = p_b + p_a$, combining the results in Lemmas~\ref{Lem:Pa} and \ref{Lem:Pb:Markov} leads to the following first main result as follows. 

\begin{proposition}\label{Prop:OnHar:CITX}Consider  the scenario where BSs adopt  channel-independent  transmission and are  powered by on-site harvesters.  The outage probability can be bounded as 
\emph{
\begin{equation}
\Pout^{\text{id}}\leq \frac{2}{2+\alpha \pi\psi}\!\! \l(\frac{c_2\theta\E\l[H_0^{-1}\r]\lambda_u}{\lambda_b^{1+\frac{\alpha}{2}}\gamma\eta}\r)^{\min(\pi\psi, 1)}\!\!+\frac{c_3\theta\lambda_u\E\l[H_0^{-1}\r]}{\lambda_b^{1+\frac{\alpha}{2}}\eta\gamma}. \nn
\end{equation}
}
\end{proposition}

Consider the scenario where the maximum harvested power  $\gamma\eta$ is large and the characteristic parameter $\psi$ is small. Then the outage probability decreases monotonically with these parameters following the scaling law of 
\begin{equation}
\Pout^{\text{id}}\propto \l(\gamma\eta\r)^{-\pi\psi}
\end{equation}
which is verified by simulation in the sequel.

Next, $p_b$ is analyzed using large deviation theory. To this end, it is necessary to  consider a specific type of distribution for the channel coefficient  $H_0$. Let $\bar{F}$ denote the \emph{complementary cumulative distribution function} (CCDF) of the random variable (RV) $H_0^{-1}$ in \eqref{Eq:p_b}. Assume that $H_0^{-1}$ is  a \emph{regularly varying} random variable defined by the following condition: 
\begin{equation}
\lim_{t\rightarrow\infty}\frac{\bar{F}(\zeta t)}{\bar{F}(t)} =\zeta^{-\omega}
\end{equation}
where $\omega > 0$ is the exponent of the distribution  and $\zeta > 1$. One example is  that $H_0$ has the chi-squared distribution that  is typical for wireless channels: 
\begin{equation}\label{Eq:Chi2}
\Pr(H_0 < t)  = \frac{1}{\Gamma(\omega)}\int^t_0 x^{\omega - 1}e^{-x}dx, \qquad t \geq 0
\end{equation}
where  $\omega > 0$ is a positive integer and the distribution parameter   and  $\Gamma$ denotes the gamma function. Then $H_0^{-1}$ is a regularly varying RV with the exponent $\omega$. Based on the assumption and applying Breiman's Theorem (see Lemma~\ref{Lem:Breiman} in Appendix~\ref{Lem:Pb:LD:Proof}), an asymptotic bound on $p_b$ is obtained as shown below. 

\begin{lemma}\label{Lem:Pb:LD}Suppose that the inverse of a channel coefficient is a regularly varying RV.  As  $\eta\gamma \rightarrow \infty$, the probability $p_b$ is bounded as\footnote{Two asymptotic relation operators, ''$\sim$" and ''$\preceq$", are defined as follows. Two functions $h(x)$ and $q(x)$ are \emph{asymptotically equivalent}, denoted as $h(x) \sim g(x)$, if $\lim_{x\rightarrow\infty}\frac{h(x)}{g(x)}=1$. The case of 
$\lim_{x\rightarrow\infty}\frac{h(x)}{q(x)}\leq 1$ is represented by $h(x)\preceq q(x)$. }
\begin{equation}\nn
p_b \preceq \E[K_0^\omega]\Pr\l(H_0 \leq \frac{\l(\frac{2\lambda_b}{3\sqrt{3}}\r)^{\frac{\alpha}{2}}\theta}{\eta\gamma}\r), \qquad  \eta\gamma \rightarrow\infty
\end{equation}
where $\E[K_0^\omega]$ is given as
\begin{equation}\label{Eq:PoissonMom}
\E[K_0^\omega] = \sum_{m=1}^\omega \l(\frac{\lambda_u}{\lambda_b}\r)^m \frac{1}{m\!}\sum_{k=0}^m(-1)^{m-k} \binom{m}{k}k^\omega. 
\end{equation}
In particular, if a channel coefficient is a chi-squared RV with parameter $\omega$, 
\begin{equation}\nn
p_b \preceq \E[K_0^\omega] \l(\frac{\eta\gamma}{\l(\frac{2\lambda_b}{3\sqrt{3}}\r)^{\frac{\alpha}{2}}\theta}\r)^{-\omega} + O\l((\eta\gamma)^{-\omega-1}\r), \quad \eta\gamma \rightarrow\infty. 
\end{equation}
\end{lemma}

\begin{remark} It is possible to derive asymptotic upper bounds on $p_b$ for other types of channel coefficient distribution. For instance, \cite[Theorem~$2.1$]{Cline:SubexpProductRVs:1994} can be applied to obtain a bound for the \emph{sub-exponential} distribution defined by the condition $\lim_{x\rightarrow\infty}\frac{\overline{F*F}(x)}{\bar{F}(x)} = 2$ where `*' denotes convolution. Such results will change the second but not the first term of the upper bound on $\Pout^{\text{id}}$ in Proposition~\ref{Prop:OnHar:CITX:LD} presented shortly. 
\end{remark}

Combining the results in Lemmas~\ref{Lem:Pa} and \ref{Lem:Pb:LD} leads to the second main result as follows: 

\begin{proposition}\label{Prop:OnHar:CITX:LD}Consider the scenario where BSs adopt channel-independent transmission and are powered by on-site harvesters. Suppose that the inverse of a channel coefficient is a regularly varying RV, the outage probability can be bounded as $\eta\gamma\rightarrow\infty$ as follows: 
\emph{
\begin{align}
\Pout^{\text{id}}&\preceq \frac{2}{2+\alpha \pi\psi} \l(\frac{\lambda_b^{1+\frac{\alpha}{2}}\gamma\eta}{c_2\theta\E\l[H_0^{-1}\r]\lambda_u}\r)^{-\min(\pi\psi, 1)} + \nn\\
& \qquad \E\l[K_0^\omega\r]\l(\l(\frac{2\lambda_b}{3\sqrt{3}}\r)^{\frac{\alpha}{2}}\theta\r)^{\omega}\Pr\l(H_0< \frac{1}{\eta\gamma}\r). \nn
\end{align}
}
In particular, if channel coefficients follow i.i.d. chi-squared distributions, as $\eta\gamma\rightarrow\infty$, 
\emph{
\begin{equation}
\begin{aligned}
&\Pout^{\text{id}}\preceq \frac{2}{2+\alpha \pi\psi} \l(\frac{\lambda_b^{1+\frac{\alpha}{2}}\gamma\eta}{c_2\theta\Gamma(\omega-1)\lambda_u}\r)^{-\min(\pi\psi, 1)}+\nn\\
&\qquad  \qquad \frac{\E[K_0^\omega]}{\Gamma(\omega+1)} \l(\frac{\eta\gamma}{\l(\frac{2\lambda_b}{3\sqrt{3}}\r)^{\frac{\alpha}{2}}\theta}\r)^{-\omega} + O\l((\eta\gamma)^{-\omega-1}\r). 
\end{aligned}\label{Prop:OnHar:CITX:LD:a}
\end{equation}
}
\end{proposition} 

Comparing the results in Propositions~\ref{Prop:OnHar:CITX} and \ref{Prop:OnHar:CITX:LD}, both upper bounds on $\Pout^{\text{id}}$ have identical first term, which is dominant when $\eta\gamma$ is large. The second term in \eqref{Prop:OnHar:CITX:LD:a}, proportional to $(\gamma\eta)^{-\omega}$, decays faster than the counterpart in   Proposition~\ref{Prop:OnHar:CITX}, proportional to $(\gamma\eta)^{-1}$, when  $\omega > 1$. This is due to a tighter bound on $p_b$ derived using large deviation theory with respect to that obtained using Markov's inequality. 

\subsection{Network Coverage with Channel Inversion Transmission}

Following the same method as in the preceding subsection, the outage probability in \eqref{Eq:Pout:UA}  can be decomposed as $\Pout^{\text{iv}} \leq  p_c+p_d$ where 
\begin{align}
p_c &= \Pr\l(P_0< \theta  \sum_{n=1}^{K_0} \frac{R_{0, n}^\alpha}{H_{0, n}} \l| \sum_{n=1}^{K_0} \frac{R_{0, n}^\alpha}{H_{0, n}}\leq \frac{\eta\gamma}{\theta} \r.\r)\times \nn\\
&\qquad \Pr\l(\sum_{n=1}^{K_0} \frac{R_{0, n}^\alpha}{H_{0, n}}\leq \frac{\eta\gamma}{\theta} \r),\label{Eq:Pd}\\
p_d &= \Pr\l(\sum_{n=1}^{K_0} \frac{R_{0, n}^\alpha}{H_{0, n}}> \frac{\eta\gamma}{\theta} \r).\label{Eq:Pc}
\end{align}

Their key difference from those of their  counterparts $p_b$ and $p_a$ in the preceding section is that the transmission power of a typical BS for the current case is a compound Poisson random variable. However, the expected transmission powers for both cases are identical due to the following equality: 
\begin{equation}\label{Eq:Cmp:Poisson:Exp}
\E\l[\sum_{n=1}^{K_0} \frac{R_{0, n}^\alpha}{H_{0, n}}\r] = \E[K_0] \E\l[R_{0, n}^\alpha\r]\E\l[H_{0, n}^{-1}\r]. 
\end{equation}
Given this equality, the bounds on $p_a$ in \eqref{Eq:p_a} and $p_b$ in \eqref{Eq:p_b} can be shown to hold for $p_c$ in \eqref{Eq:Pd} and $p_d$  in \eqref{Eq:Pc}, respectively, yielding the following proposition. 

\begin{proposition}\label{Prop:OnSite:UATX}Consider  the scenario of  on-site harvesters. The upper bound on the outage probability for the case of channel-independent transmission as given in Proposition~\ref{Prop:OnHar:CITX} also holds for  the case of channel inversion transmission. 
\end{proposition}

Despite the identical upper bounds, the outage probability for channel inversion transmission is lower than that for channel-independent transmission. The performance gain is due to  adapting transmission-power allocation to multiuser channel states to minimize the number of mobiles in outage. 

Next, $p_c$ is analyzed using large deviation theory. To this end, some useful definitions and results are introduced as follows. 
The distribution of a RV $X$ has a \emph{light tail} if $\E\l[e^{\delta X}\r] < \infty$ for some $\delta > 0$ and a \emph{heavy tail} if  $\E\l[e^{\delta X}\r] =\infty$ for all  $\delta > 0$. The following result is from \cite[Theorem~$2$]{Denisov:LowerLimRandomlyStoppeSums:2008}. 
\begin{lemma}\label{Lem:RandSum}Let $\{X_m\}$ be a set of i.i.d. random variables having  a heavy-tailed distribution and $N$ have a light-tailed distribution independent of  those of $\{X_m\}$. Then 
\begin{equation}\nn
\Pr\l(\sum_{m=1}^N X_m > t\r) \sim \E[N] \Pr(X_1 > t)\qquad \text{as} \ t\rightarrow\infty. 
\end{equation}
\end{lemma}
To apply the result,  the upper bound on $p_d$ in  \eqref{Eq:Pc} is simplified using  the inequality $R_{0, n} \leq \sqrt{2\lambda_b/3\sqrt{3}}$ as
\begin{equation}
p_d \leq \Pr\l(\sum_{n=1}^{K_0} H_{0, n}^{-1}> \frac{\eta\gamma}{\l(2\lambda_b/3\sqrt{3}\r)^{\frac{\alpha}{2}}\theta} \r). \label{Eq:Pd:UB}
\end{equation}
One can see that the upper bound is the tail probability of a compound Poisson RV,  $\sum_{n=1}^{K_0} H_{0, n}^{-1}$. Note that the Poisson distribution of $K_0$ has a light tail since it decays faster than the exponential rate as observed from the following inequality (see e.g., \cite[Theorem~5.4]{MitzenmacherBook:RandomAlgo:05}):
\begin{equation}
\Pr(K_0  \geq x)\leq \frac{e^{-\lambda_m/\lambda_b}(e\lambda_m/\lambda_b)^x}{x^x}, \qquad \text{if}\ x > \lambda_m/\lambda_b. \nn
\end{equation}
Then the lemma below follows from \eqref{Eq:Pd:UB} and Lemma~\ref{Lem:RandSum}, which is the channel inversion counterpart of Lemma~ \ref{Lem:Pb:LD}. 

\begin{lemma}\label{Lem:Pd:LD}Suppose that the inverse of a channel coefficient has a heavy-tailed distribution.  As  $\eta\gamma \rightarrow \infty$, the probability $p_d$ is bounded as 
\begin{equation}\label{Eq:Pd:LD:a}
p_d \preceq \frac{\lambda_u}{\lambda_b}\Pr\l(H_0 \leq \frac{\l(\frac{2\lambda_b}{3\sqrt{3}}\r)^{\frac{\alpha}{2}}\theta}{\eta\gamma}\r), \qquad  \eta\gamma \rightarrow\infty
\end{equation}
In particular, if a channel coefficient is a chi-squared RV with parameter $\omega$, 
\begin{equation}\nn
p_d \preceq \frac{\lambda_u}{\lambda_b} \l(\frac{\eta\gamma}{\l(\frac{2\lambda_b}{3\sqrt{3}}\r)^{\frac{\alpha}{2}}\theta}\r)^{-\omega} + O\l((\eta\gamma)^{-\omega-1}\r), \qquad  \eta\gamma \rightarrow\infty. 
\end{equation}
\end{lemma}

As mentioned, the bound on $p_a$ in \eqref{Eq:p_a} holds for  $p_c$ in \eqref{Eq:Pd}. Combining this result, the bounds on  $p_d$  in Lemma~\ref{Lem:Pd:LD} and the inequality $\Pout^{\text{iv}} \leq  p_c+p_d$ gives the following result. 

\begin{proposition}\label{Prop:OnHar:INVTX:LD}Consider  the scenario where BSs adopt channel inversion transmission and are  powered by on-site harvesters. Suppose that the inverse of a channel coefficient is a heavy-tailed  RV, the outage probability can be bounded as follows: as $\eta\gamma\rightarrow\infty$, 
\begin{align}
\Pout^{\text{iv}}&\preceq \frac{2}{2+\alpha \pi\psi} \l(\frac{\lambda_b^{1+\frac{\alpha}{2}}\gamma\eta}{c_2\theta\E\l[H_0^{-1}\r]\lambda_u}\r)^{-\min(\pi\psi, 1)}+\nn \\
&\qquad \frac{\lambda_u}{\lambda_b} \Pr\l(H_0< \l(\frac{2\lambda_b}{3\sqrt{3}}\r)^{\frac{\alpha}{2}}\frac{\theta}{\eta\gamma}\r). \nn
\end{align}
In particular, if channel coefficients follow i.i.d. chi-squared distributions, as $\eta\gamma\rightarrow\infty$, 
\begin{equation}
\begin{aligned}
\Pout^{\text{iv}}&\preceq \frac{2}{2+\alpha \pi\psi} \l(\frac{\lambda_b^{1+\frac{\alpha}{2}}\gamma\eta}{c_2\theta\Gamma(\omega-1)\lambda_u}\r)^{-\min(\pi\psi, 1)}+\\
&\qquad  \frac{\lambda_u}{\Gamma(\omega+1)\lambda_b} \l(\frac{\eta\gamma}{\l(\frac{2\lambda_b}{3\sqrt{3}}\r)^{\frac{\alpha}{2}}\theta}\r)^{-\omega} + O\l((\eta\gamma)^{-\omega-1}\r).
\end{aligned}\nn
\end{equation}
\end{proposition} 

Comparing the results in Propositions~\ref{Prop:OnHar:CITX:LD} and \ref{Prop:OnHar:INVTX:LD}, the outage probability for channel inversion transmission follows the same scaling laws as those  for channel-independent transmission except for some difference in linear factors. As a result, $\Pout^{\text{iv}}$ is smaller than $\Pout^{\text{id}}$, as also observed in the simulation results in the sequel.

\section{Network Coverage with Distributed Harvesters}\label{Section:Coverage:DistHar}
\subsection{Network Coverage}
In this section, it is shown that aggregating energy harvested by many distributed harvesters stabilizes the power supplied to the BSs by the law of large numbers. It is assumed that the energy aggregation loss is regulated such that the scaling factor of BS transmission powers due to such loss is smaller than a constant $\tau \in (0, 1)$. The design requirement under this constraint is analyzed in the next subsection. The harvester lattice partitions the plane into small hexagonal regions, each  with area of $1/\lambda_h$. 
Whether each region contains an energy center can be indicated by a set of independent Bernoulli random variables $\{Q_n\}$  with probabilities $\exp(-\lambda_e/\lambda_h)$ and $[1 - \exp(-\lambda_e/\lambda_h)]$ for the values of $0$ and $1$, respectively. 
Despite the independence of the numbers of energy center in different regions, it is important to note that the energy intensities at  harvesters are correlated [see their joint distribution in \eqref{Eq:JCDF} and \eqref{Eq:JCDF:a}]. The energy intensity at each harvester is at least $\exp\l(-\frac{2}{3\sqrt{3}\nu\lambda_h}\r)$ if the corresponding small region contains an energy center or otherwise  takes on some positive value. Based on the above points, the transmission power of the typical BS is lower bounded as
\begin{align}
P_0 &\geq \frac{\tau \gamma\eta\lambda_a}{\lambda_b} \sum_{n=1}^{M_0} Q_n e^{\l(-\frac{2}{3\sqrt{3}\nu\lambda_h}\r)}\nn\\
&= \frac{\tau \gamma\eta\lambda_h}{\lambda_b}\times\frac{\lambda_a M_0}{\lambda_h}\times \frac{1}{M_0}\sum_{n=1}^{M_0} Q_n e^{-\frac{2}{3\sqrt{3}\nu\lambda_h}}   \label{Eq:Average}
\end{align}
where $M_0$ represents the number of harvesters connected to the typical aggregator. Consider the scenario of sparse aggregators with each connected to many harvesters, corresponding to $\lambda_a\rightarrow 0$. As a result, $M_0\rightarrow \infty$ and $\lambda_a M_0/\lambda_h\rightarrow 1$. Then it follows from \eqref{Eq:Average} that 
\begin{equation}
\!\lim_{\lambda_a\rightarrow 0}P_0(\lambda_a) \!\geq\! \frac{\tau\gamma\eta\lambda_h}{\lambda_b} \lim_{M_0\rightarrow \infty}\frac{1}{M_0}\sum_{n=1}^{M_0} Q_n e^{\l(-\frac{2}{3\sqrt{3}\nu\lambda_h}\r)}. \label{Eq:Average:a}
\end{equation}
The law of large numbers gives the following lemma. 
\begin{lemma}\label{Lem:EnAgg}With the  harvester density $\lambda_h$ fixed, as the aggregator density $\lambda_a\rightarrow 0$, the transmission power for the typical BS converges as 
\begin{equation}\label{Eq:P:Lim}
\lim_{\lambda_a\rightarrow 0}P_0(\lambda_a) \geq \frac{\tau\gamma\eta\lambda_h}{\lambda_b} \l(1 - e^{-\frac{\lambda_e}{\lambda_h}}\r) e^{-\frac{2}{3\sqrt{3}\nu\lambda_h}}, \quad \text{a.s.}  
\end{equation}
\end{lemma}
In other words, $P_0$ is lower bounded by a constant and thus its randomness due to energy spatial variation diminishes.  If the energy centers are dense ($\lambda_e \gg \lambda_h$) and the shape parameter $\nu$ is large  ($\nu\lambda_h \gg 1$), the transmission power approaches its upper bound $\gamma\eta\lambda_h/\lambda_b$. 

The power stabilization by the spatial averaging of the energy field  removes one random variable from the outage probability. Consider the case of channel-independent transmission and the corresponding outage probability in \eqref{Eq:Pout:CI}. Using Lemma~\ref{Lem:EnAgg} and applying Markov's inequality, the outage probability for small aggregator density can be bounded as 
\begin{equation}
\lim_{\lambda_a\rightarrow 0}\Pout^{\text{id}}(\lambda_a) \leq \frac{\theta\lambda_b\E[H^{-1}]\E[K]\E[R^\alpha]}{\tau\gamma\eta\lambda_h \l(1 - e^{-\frac{\lambda_e}{\lambda_h}}\r) e^{-\frac{2}{3\sqrt{3}\nu\lambda_h}}}. 
\end{equation}
Substituting $\E[K] = \lambda_u/\lambda_b$ and the result in \eqref{Eq:R0:alp} yields the result in Proposition~\ref{Prop:Pout:Dist} as given below. The result is proved to also hold for channel-inversion transmission  using \eqref{Eq:Pout:UA} and following a similar procedure. 
\begin{proposition}\label{Prop:Pout:Dist}With the  harvester density $\lambda_h$ fixed, as the aggregator density $\lambda_a\rightarrow 0$, the outage probabilities for both the channel-independent and channel inversion transmissions can be bounded as
\begin{equation}
\lim_{\lambda_a\rightarrow 0}\Pout^{\text{id}}(\!\lambda_a\!), \lim_{\lambda_a\rightarrow 0}\Pout^{\text{iv}}(\!\lambda_a\!) \!\leq\! \frac{c_3\theta\E\l[H^{-1}\r]\lambda_u}{\tau\gamma\eta \lambda_b^{\frac{\alpha}{2}}\lambda_h\!\!\l(\!1\! -\! e^{-\frac{\lambda_e}{\lambda_h}}\!\r)\! e^{-\frac{2}{3\sqrt{3}\nu\lambda_h}}}. \nn\\
\end{equation}
\end{proposition}
The results in Proposition~\ref{Prop:Pout:Dist} suggest that 
\begin{equation}
\lim_{\lambda_a\rightarrow 0}\Pout^{\text{id}}(\lambda_a), \lim_{\lambda_a\rightarrow 0}\Pout^{\text{iv}}(\lambda_a) \propto \frac{1}{\gamma\eta}\times \frac{1}{\lambda_h/\lambda_b}\times \frac{\lambda_u}{\lambda_b}\times \lambda_b^{-\frac{\alpha}{2}}\nn
\end{equation}
where the factors represent in order the inverses of the maximum power generated by a single harvester, the number of harvesters per BS, the expected number of active  mobiles õper cell,  and the expected propagation loss. 

\subsection{Energy Aggregation  Loss}
It is well known that the power line loss, $\Delta P$, for transmitting the power of $P$ to a receiver is given as 
\begin{equation}
\Delta P = \frac{\beta P^2d}{V^2} \label{Eq:Delta:P} 
\end{equation}
where $V$ is the  voltage and $\beta$ is a constant depending on the power line characteristics, such as resistivity and cross-section area. In other words, the total transmission power is $(P + \Delta P)$. Consider an arbitrary harvester in a typical cluster. Let $V_0$, $g_0$, and $d_0$ denote the transmission voltage and  energy intensity at this harvester and the corresponding distance for power transmission to the connected aggregator, respectively. Then the power harvested at and transmitted by the harvester is $\eta g_0$. By the definition of transmission efficiency $\tau$, the  transmission loss, denoted as $\Delta P_0$, should be no larger than the  fraction $(1-\tau)$ of transmission power: 
\begin{equation}
\Delta P_0 \leq (1-\tau) \eta g_0. \label{Eq:AggLoss:b}
\end{equation}
Since the power effectively transmitted from the harvester to the aggregator is $\tau \eta g_0$, it follows from \eqref{Eq:Delta:P} that 
\begin{equation}
\Delta P_0 = \frac{\beta (\tau \eta  g_0)^2}{V_0^2} d_0.  \label{Eq:AggLoss:a}
\end{equation}
Combining \eqref{Eq:AggLoss:b} and \eqref{Eq:AggLoss:a} leads to a requirement on the voltage
\begin{equation}
V_0 \geq \tau \sqrt{\frac{\beta \eta g_0 d_0}{1-\tau}}. \label{Eq:Volt:Req}
\end{equation}
The distance $d_0$ can be bounded as $d_0 \leq \sqrt{\frac{2}{3\sqrt{3}\lambda_a}}$ since the harvester lies in a hexagon centered at the connected aggregator and with an area of $1/\lambda_a$. Using this bound as well as $g_0 \leq \gamma$, a sufficient condition for meeting the requirement in \eqref{Eq:Volt:Req} is obtained as shown in the following proposition. 

\begin{proposition}\label{Prop:Voltage} Consider distributed harvesters and aggregators with fixed densities. A sufficient condition on the harvester voltage for achieving a given transfer efficiency $\tau \in (0, 1)$ is 
\begin{equation}
V_0 = \tau\sqrt{\frac{\beta \eta \gamma}{1-\tau}\sqrt{\frac{2}{3\sqrt{3}}}}\lambda_a^{-\frac{1}{4}}. \label{Eq:Voltage}
\end{equation}
\end{proposition}
Given  fixed $\tau$,   one can observe from  Proposition~\ref{Prop:Voltage} that as the size of  harvester clusters increases by letting $\lambda_a \rightarrow 0$, it is sufficient for the harvester voltage to scale as $\lambda_a^{-1/4}$ multiplied by a constant, resulting in  reliable power supply to all BSs. 

It is also interesting to consider the case with fixed $V_0$ for which there exists a conflict between suppressing energy spatial randomness by energy aggregation  and reducing  the resultant  power-transmission loss. Specifically, based on 
\eqref{Eq:Voltage}, decreasing the aggregator density $\lambda_a$ reduces the transmission efficiency $\tau$, which in turn increases the outage-probability upper bound in Proposition~\ref{Prop:Pout:Dist}.  This makes it necessary to optimize $\lambda_a$ for  minimizing  the outage probability. Solving the problem is non-trivial and outside the scope of this paper but warrants further investigation.

\section{Extensions and Discussion}\label{Section:Ext}

\subsection{Shot-Noise Based  Energy Field Model}\label{App:Field:Alt}
In this subsection, the energy field model proposed in Section~\ref{Section:Field} is compared with an alternative model described as follows.  By  replacing the max operator in the energy intensity  function in \eqref{Eq:EnDen} with a summation, the result,  denoted as $\tilde{g}(X)$,  represents spatial interpolation of the energy centers:
\begin{equation}\label{Eq:ShotNoise}
\tilde{g}(X) =\gamma\sum _{Y \in \Phi_e} f(|X-Y|). 
\end{equation}
One drawback of the resultant alternative model is that the maximum energy intensity of the field can no longer be retained as $\gamma$ as it becomes a random variable with unbounded support, which is impractical.  As another drawback, the random function in \eqref{Eq:ShotNoise} is known as a \emph{shot-noise process} and its  distribution function has no simple form \cite{Lowen:PowerLawShotNoise:1990}.

The expectation of the energy density in the alternative model in \eqref{Eq:ShotNoise} can be obtained using  Campbell's theorem \cite{Kingman93:PoissonProc}: 
\begin{align}
\E[\tilde{g}(X)]  &= \gamma\lambda_e \int_0^\infty e^{-\frac{r^2}{\nu}} 2\pi r dr\nn\\
&= \pi \gamma \psi. 
\end{align}
Hence, the expectation ratio $\E[\tilde{g}(X)]/\E[g(X)] = 1 + \pi \psi$. As $\psi$ is typically much smaller than $1$, the ratio is close to one.  

Using the result and the technique  in \cite{WeberAndrews:TransCapWlssAdHocNetwkSIC:2005} for bounding the tail probability of a shot-noise process, it can be obtained that 
\begin{align}
\Pr(\tilde{g}(X) > x) &= \Pr(g(X) > x) +  \Pr(g(X) \leq x) \times \nn\\
&\qquad \Pr\Biggl(\sum _{Y \in \Phi\cap \bar{B}(X, f^{-1}(x/\gamma))} f(|X-Y|) > \frac{x}{\gamma}\Biggr)\nn\\
& \leq  \Pr(g(X) > x) +  \Pr(g(X) \leq x)\times \nn\\
&\qquad \frac{\E\l[\sum _{Y \in \Phi\cap \bar{B}(X, f^{-1}(x/\gamma))} f(|X-Y|)\r]}{x/\gamma}\nn\\
&= \Pr(g(X) > x) +  \Pr(g(X) \leq x) \times \pi \lambda_e \nu. \nn
\end{align}
The inequality can be rewritten as 
\begin{align}
\Pr(\tilde{g}(X) > x) - \Pr(g(X) > x) &\leq  \pi \lambda_e \nu\times \Pr(g(X) \leq x)\nn\\
&\leq \pi \lambda_e \nu\epsilon \label{Eq:g:compare}
\end{align}
where  the constant $\epsilon\in (0, 1)$ is related to the maximum outage probability of the renewable powered network and  is close to zero in the operating regime of interest. It follows that the tail probabilities for both energy field models are similar. 

\subsection{Energy Field Model  with Power Law Energy Decay and Its Effect on Network Coverage}\label{Eq:Field:Ext}
In the preceding sections, the outage probability is analyzed based on the energy field model characterized by the exponential energy decay function of distance in \eqref{Eq:DecayFun}. In this section, the analysis is extended to the alternative power law function $f'$ defined  as
\begin{equation}\label{Eq:Decay:Pwr}
f'(d) = \l(1 + \frac{d^2}{\nu}\r)^{-1}
\end{equation}
where $f'(d) \propto d^{-2}$ if $d$ is large. Note that $f'(d) \rightarrow 0$ as $d \rightarrow \infty$ and $f'(d) \rightarrow 1$ as $d\rightarrow 0$, having the desirable properties for an energy decay function. It is assumed that the BSs adopt channel-independent transmission and that the channel coefficients follow a chi-squared distribution with parameter $\omega$. A similar procedure can be followed to obtain results for other cases, however the key insights are identical to those from the current analysis. Therefore, the details are omitted for brevity. 

Recall that the outage probability can be decomposed as $\Pout^{\text{id}} = p_a + p_b$ with $p_b$ and $p_a$ defined in \eqref{Eq:p_a} and  \eqref{Eq:p_b}. From their definitions, the change on the energy field model by modifying the energy decay function affects only $p_a$ but not $p_b$. The resultant $p_a$ is analyzed as follows. By a slight abuse of notation, let the distance function $r(\cdot)$ in \eqref{Eq:Radius} and  the energy intensity function $g(\cdot)$ in \eqref{Eq:EnDen} also denote their counterparts corresponding to  $f'(\cdot)$ in \eqref{Eq:Decay:Pwr}. Then 
\begin{equation}
r(x) =f'^{-1}\l(x/\gamma\r) = \sqrt{\nu\l(\gamma/x-1\r)}. 
\end{equation}
This results in the following distribution of the  energy intensity  $g(X)$ for an arbitrary location $X\in\mathds{R}^2$:
\begin{align}
\Pr(g(X) \leq x) &= e^{-\pi \lambda_e r^2(x)}\nn\\
&= e^{-\pi \psi \l(\frac{\gamma}{x} - 1\r)}. \nn
\end{align}
Then from \eqref{Eq:p_a}, $p_a$ can be upper bounded as 
\begin{align}
p_a &\leq \Pr\l(\frac{g(X) H_0 R_0^{-\alpha}}{K_0}   \leq \frac{\theta}{\eta}\r) \nn\\
&= e^{\pi\psi}\E\l[\exp\l(-\frac{\pi\psi\gamma\eta H_0}{\theta K_0R_0^\alpha}\r)\r]\nn\\
&= \l(\frac{\theta }{\pi\psi\eta\gamma}\r)^{\omega}e^{\pi\psi} \E\l[K_0^\omega\r]\E\l[R_0^{\alpha \omega}\r]\nn\\
&= \l(\frac{\theta }{\pi\psi\eta\gamma}\r)^{\omega}e^{\pi\psi} \l(\frac{2}{3\sqrt{3}\lambda_b}\r)^{\frac{\alpha\omega}{2}}\E\l[K_0^\omega\r]\label{Eq:Pa:UB:New}
\end{align}
where $\E\l[K_0^\omega\r]$ is given in \eqref{Eq:PoissonMom}. Combining the last inequality with Lemmas~\ref{Lem:Pb:Markov} and \ref{Lem:Pb:LD} gives the main result of this section. 

\begin{proposition}\label{Prop:OnHar:CITX:New}Consider the scenario where BSs adopt channel-independent  transmission and are  powered by on-site harvesters. Suppose that the channel coefficients follow i.i.d. chi-squared distributions and that the energy density decay follows the function $f'$ in \eqref{Eq:Decay:Pwr}, the outage probability satisfies
\begin{equation}
\Pout^{\text{id}}\leq \l(\frac{\theta }{\pi\psi\eta\gamma}\r)^{\omega}e^{\pi\psi} \l(\frac{2}{3\sqrt{3}\lambda_b}\r)^{\frac{\alpha\omega}{2}}\E\l[K_0^\omega\r] +\frac{c_3\theta\lambda_u\E\l[H_0^{-1}\r]}{\lambda_b^{1+\frac{\alpha}{2}}\eta\gamma}. \nn
\end{equation}
Moreover, as $\eta\gamma\rightarrow\infty$, 
\begin{equation}
\Pout^{\text{id}}\preceq c_5\l[\l(\frac{1}{\pi\psi}\r)^{\omega}e^{\pi\psi} + \frac{1}{\Gamma(\omega+1)}\r](\eta\gamma)^{-\omega} + O\l((\eta\gamma)^{-\omega-1}\r) \nn
\end{equation}
where the constant 
\begin{equation}
c_5 = \E[K_0^\omega]\theta^{\omega} \l(\frac{2\lambda_b}{3\sqrt{3}}\r)^{\frac{\omega\alpha}{2}}.  \nn
\end{equation}
\end{proposition}

Recall that the energy field influences the outage probability via its spatial variation and maximum harvested power, determined by the parameters $\psi$ and $\eta\gamma$, respectively. The results show that the component of the outage probability due to energy randomness  scales as  $\psi^{-\omega}$ if $\psi$ is small. Moreover, as $\eta\gamma$ increases, the probability decreases following the power law $(\eta\gamma)^{-\omega}$ as observed from Proposition~\eqref{Prop:OnHar:CITX:New}.   The scaling law $(\eta\gamma)^{-1}$ is more gradual due to a loose bound resulting from  the use of Markov inequality in the derivation. 

Comparing the results in Proposition~\ref{Prop:OnHar:CITX:New} with those in Proposition~\ref{Prop:OnHar:CITX:LD} corresponding to the exponential energy decay function, the outage probability for the current case is less sensitive to the changes on $\psi$ but more sensitive to those on $\eta\gamma$. The reason is that the power law energy decay function results in less spatial fluctuation in the energy field with respect to the double-exponential function.

\begin{figure}[t]
\begin{center}
\includegraphics[width= 8.5 cm]{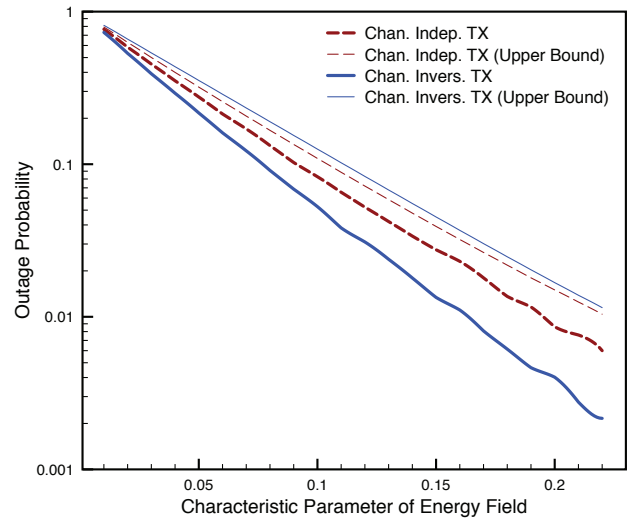}
\caption{Outage probability  versus the characteristic parameter of the energy field for the scenario of on-site harvesters.}
\label{Fig:OnHar}
\end{center}
\end{figure}

\section{Simulation Results}\label{Section:Sim}

Unless otherwise specified, the simulation settings are as follows. The BSs and mobiles have densities of $\lambda_b = 0.78  /\text{km}^2$ and $\lambda_m = 7.8  /\text{km}^2$, respectively, corresponding to an average cell radius of  $500$ meter and $10$ users per cell. For propagation, the reference path loss is $70$ dB measured for a propagation distance of $100$m \cite{rappaport}, the pathloss exponent $\alpha = 4$, and the noise power is $-90$ dBm. The product of the  harvester aperture and the maximum energy intensity,  $\gamma\eta$, gives the maximum power a harvester can generate, which is fixed as  $\gamma\eta = 1 $ kW for an on-site harvester and $10$ W for a distributed one. For the case of distributed harvesters, the harvester density is $15.6  /\text{km}^2$  that is twice of the BS density.  The   SNR threshold $\theta$ is $8$ corresponding to a spectrum efficiency  of about $3.2$ bit/s/Hz. The fading coefficients are i.i.d. and distributed as  $\max(|\mathcal{CN}(1, 1)|^2, 0.1)$,  where the $\mathcal{CN}(1, 1)$ random variables model Rician fading and the truncation at $0.1$ accounts for the avoidance of deep fading by scheduling. 

Consider the scenario of on-site harvesters.  The curves of outage probability  versus the characteristic parameter $\psi$  are plotted in Fig.~\ref{Fig:OnHar}  for both the channel-independent and channel inversion transmissions. The upper bounds on the outage probabilities as given in Propositions~\ref{Prop:OnHar:CITX} and \ref{Prop:OnSite:UATX} are also shown.  The outage probabilities are observed to decay exponentially with increasing $\psi$ as predicted by analysis. Channel inversion transmission perform better than the other transmission scheme. The probability bound for channel inversion transmission is not as tight as that for channel-independent transmission as the former is based on the union bound [see  \eqref{Eq:Pout:UA}]. As the product $\gamma\eta$ is large, significant power can be harvested even at locations far from energy centers. Consequently,  the outage probability is close to zero even for a small characteristic parameter (e.g., $0.2$).

Next, consider the scenario of distributed harvesters  where the characteristic parameter is fixed as $0.05$.  In Fig.~\ref{Fig:DistHar}, the outage probability is plotted against the number of harvesters  connected to a single aggregator for energy aggregation (that is approximately equal to $\lambda_h/\lambda_a$).  For comparison, the figure also shows the asymptotic upper bound on outage probabilities as given in Proposition~\ref{Prop:Pout:Dist} as well as those generated by simulation and replacing transmission powers with the asymptotic lower bound in Lemma~\ref{Lem:EnAgg}.  Aggregation loss is omitted assuming sufficiently high transmission voltage.  It is observed that energy aggregation dramatically reduces outage probabilities, indicating energy randomness as the main reason  for outage events.  Most of the aggregation gain can be achieved with less than $50$ harvesters per aggregator. For a large number of harvesters, the limits of outage probability depend only slightly   on the randomness of the energy field and is mostly affected  by channel fading as well as mobile random locations for the case of channel inversion transmission.  Last, the asymptotic upper bound from Proposition~\ref{Prop:Pout:Dist} is observed to be loose due to the use of Markov's inequality but the other asymptotic bounds based on Lemma~\ref{Lem:EnAgg} are tight.

\section{Conclusion}\label{Section:Conclusion}
In this paper, a novel spatial model of renewable energy has been presented and applied to quantify the effect of energy spatial randomness on the coverage of cellular networks powered by energy harvesting. Moreover, the proposed technique of energy aggregation has demonstrated that new network architectures can be designed to cope with energy spatial variations. This work provides a useful analytical framework for designing large-scale energy harvesting networks. Moreover, it opens an interesting research direction of redesigning communication  techniques e.g., multi-cell cooperation and resource allocation, as a means to achieve network reliability in the presence of energy randomness. 

\appendix 

\begin{figure}[t]
\begin{center}
\includegraphics[width= 8.5 cm]{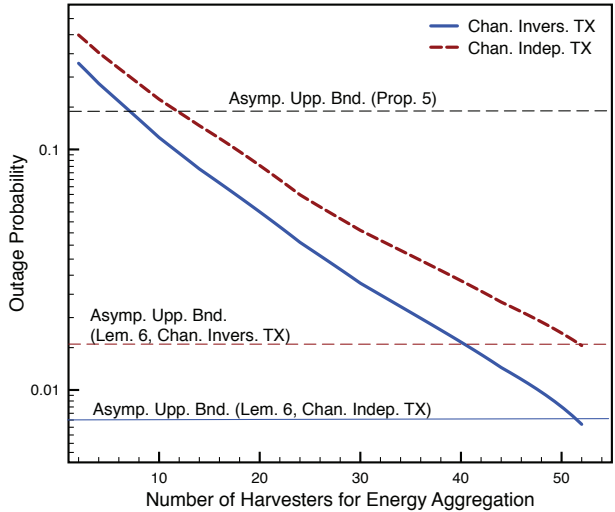}
\caption{Outage probability  versus the number of harvesters for energy aggregation for the scenario of distributed  harvesters. The characteristic parameter is fixed as $\psi = 0.05$. }
\label{Fig:DistHar}
\end{center}
\end{figure}

\subsection{Proof of Lemma~\ref{Lem:Pa}}\label{Lem:Pa:Proof}
By substituting   $P_0 = \eta g(B_0)$ and \eqref{Eq:Field:Dist} into  \eqref{Eq:p_a}, 
\begin{align}
p_a &=\l(\frac{\theta}{\gamma\eta}\r)^{\pi\psi}\E\l[\l(K_0 H_0^{-1} R_0^{\alpha}\r)^{\pi\psi}\mid K_0 H_0^{-1} R_0^{\alpha}\leq \eta\gamma\r] \times\nn\\
&\qquad \Pr\l(K_0 H_0^{-1} R_0^{\alpha}\leq \gamma\r). \nn
\end{align}
It follows that 
\begin{equation}
p_a \leq \l(\frac{\theta}{\gamma\eta}\r)^{\pi\psi}\E[K_0^{\pi\psi}]\E\l[H_0^{-\pi\psi}\r]\E\l[R_0^{\alpha \pi\psi}\r].\label{Eq:Pout:CI:a}
\end{equation}

By substituting $\bar{R}_0$ into \eqref{Eq:Pout:CI:a}, 
\begin{align}
p_a  &\leq  \l(\frac{\theta}{\gamma\eta\lambda_b^{\frac{\alpha}{2}}}\r)^{\pi\psi}\E[K^{\pi\psi}]\E\l[H_0^{-\pi\psi}\r]\E\l[\bar{R}_0^{\alpha \pi\psi}\r]. \label{Eq:Pout:CI:b}
\end{align}
It follows from  the inequality $\bar{R}_0 \preceq D$ and the distribution of $D$ in \eqref{Eq:R:Dist:Inner} that 
\begin{equation}
\E\l[\bar{R}_0^{\alpha \pi\psi}\r] \leq \frac{2 c_2^{\pi\psi}}{2+\alpha \pi\psi}
\end{equation}
where $c_2$ is defined in the lemma statement. By combining the inequality with \eqref{Eq:Pout:CI:b}, 
\begin{equation}
p_a\leq \frac{2}{2+\alpha \pi\psi} \l(\frac{c_2\theta}{\gamma\eta\lambda_b^{\frac{\alpha}{2}}}\r)^{\pi\psi}\E[K_0^{\pi\psi}]\E\l[H_0^{-\pi\psi}\r].\label{Eq:Pout:CI:c}
\end{equation}
If  $\pi\psi \leq 1$, the upper bound on the outage probability can be reduced using Jensen's inequality as
\begin{equation}
p_a\leq \frac{2}{2+\alpha \pi\psi} \l(\frac{c_2\theta \E[K_0]\E\l[H_0^{-1}\r]}{\gamma\eta\lambda_b^{\frac{\alpha}{2}}}\r)^{\pi\psi}. \label{Eq:Pb:ub:a}
\end{equation}
If $\pi\psi > 1$, since larger $\psi$ leads to more harvested energy and thus smaller outage probability,  the inequality in \eqref{Eq:Pout:CI:c} holds by setting $\pi\psi = 1$: 
\begin{align}
p_a&\leq \frac{2c_2\theta\E[K_0]\E\l[H_0^{-1}\r]}{(2+\alpha \pi\psi)\gamma\eta\lambda_b^{\frac{\alpha}{2}}}. \label{Eq:Pb:ub:b}
\end{align}
Combining \eqref{Eq:Pb:ub:a} and \eqref{Eq:Pb:ub:b} and substituting $\E[K_0] = \lambda_u/\lambda_b$ gives the desired result.

\subsection{Proof of Lemma~\ref{Lem:Pb:Markov}}\label{Lem:Pb:Markov:Proof}
Using Markov's inequality,  $p_b$ defined in \eqref{Eq:p_b} can be bounded as
\begin{align}
p_b &\leq \frac{\theta\E\l[K_0 H_0^{-1} R_0^{\alpha}\r]}{\eta\gamma}\nn\\
&= \frac{\theta\E\l[K_0\r]\E\l[H_0^{-1}\r]\E\l[R_0^{\alpha}\r]}{\eta\gamma}\label{Eq:Pa}\\
&= \frac{\theta\lambda_u\E\l[H_0^{-1}\r]\E\l[R_0^{\alpha}\r]}{\eta\gamma\lambda_b} \label{Eq:Pa:a}
\end{align}
where the equality in \eqref{Eq:Pa} holds since $K_0$,  $H_0$ and $R_0$ are independent and \eqref{Eq:Pa:a} follows by  substituting $\E[K_0] = \lambda_u/\lambda_b$.  Using the definition of $\bar{R}_0$ and the inequality in \eqref{Eq:R:Dist:Inner}, it is obtained that 
\begin{equation}
\E\l[R_0^{\alpha}\r] \leq \l(\frac{2}{3\sqrt{3}\lambda_b}\r)^{\frac{\alpha}{2}}. \label{Eq:R0:alp}
\end{equation}
Substituting \eqref{Eq:R0:alp} into \eqref{Eq:Pa:a} gives the desired result.

\subsection{Proof of Lemma~\ref{Lem:Pb:LD}}\label{Lem:Pb:LD:Proof}
The probability $p_b$ in \eqref{Eq:p_b} can be bounded as 
\begin{align}
p_b &= \Pr\l(K_0 H_0^{-1} \bar{R}_0^{\alpha}> \frac{\eta\gamma}{\lambda_b^{\frac{\alpha}{2}}\theta} \r)\nn\\ 
&\leq \Pr\l(K_0 H_0^{-1} D^{\alpha}> \frac{\eta\gamma}{\lambda_b^{\frac{\alpha}{2}}\theta} \r)\nn\\
&\leq \Pr\l(K_0 H_0^{-1} > \frac{\eta\gamma}{\l(\frac{2\lambda_b}{3\sqrt{3}}\r)^{\frac{\alpha}{2}}\theta} \r).\label{Eq:Pb:UB}
\end{align}
Thus, analyzing the scaling law of $p_b$ as the maximum harvested power $\eta\gamma$ increases is equivalent to characterizing the large deviation of the product of the two RVs $K_0$ and $H_0^{-1}$. This relies on Breiman's Theorem stated as follows \cite[Corollary~3.6]{Cline:SubexpProductRVs:1994}. 

\begin{lemma}[Breiman's Theorem]\label{Lem:Breiman}Suppose that $X$ and $Y$ are two independent non-negative RVs where $X$ is a \emph{regularly varying} RV with the exponent $\omega  > 0$ and $\E[Y^{\omega +\epsilon}]< \infty$ for some $\epsilon > 0$. Then 
\begin{equation}
\Pr(XY > t)\sim \E[Y^\omega]\Pr(X > t), \qquad t \rightarrow \infty.
\end{equation}
\end{lemma}

It is known that the moment of the Poisson RV $K_0$  exists as given below. 
Thus, $\E[K_0^{\omega+\epsilon}] < \infty$ for some $\epsilon > 0$. Given this condition and the assumption of $H_0^{-1}$ being a regularly varying RV, the first result in the lemma statement can be obtained from \eqref{Eq:Chi2}, \eqref{Eq:Pb:UB} and Lemma~\ref{Lem:Breiman}. The second result follows by substituting the distribution of $H_0$ in \eqref{Eq:Chi2}.

\bibliographystyle{ieeetr}

\end{document}